\documentclass{jfm}

\usepackage{amsmath}
\usepackage{amsfonts}
\usepackage{amssymb}
\usepackage{graphicx}
\usepackage{gensymb}
\usepackage{caption}
\usepackage{subcaption}
\usepackage{enumitem}
\usepackage{mathtools}
\usepackage{natbib}

\usepackage[capitalise]{cleveref}

\newcommand{\pder}[2] {\frac{\partial #1}{\partial #2}}
\newcommand{\ader}[2]{u#2\pder{#1}{x}+v#2\pder{#1}{y}+w#2\pder{#1}{z}}

\newcommand{\pderline}[2] {{\partial #1}/{\partial #2}}
\newcommand{\Dder}[2] {\frac{D #1}{D #2}}

\newcommand{\da}[1] {\overline{#1}}

\newcommand{\z} {\zeta}
\newcommand{\tw}[1] {\tilde{#1}}

\DeclareMathOperator{\sech} {sech}

\newcommand{\Ro}{\mathrm{Ro}}

\newcommand{\E}{\mathrm{E}}
\renewcommand{\Pr}{\mathrm{Pr}}

\DeclareMathOperator{\dint}{d\!}

\title{The modification of turbulent thermal wind balance by non-traditional effects}
\author{Matthew N. Crowe}

\shorttitle{The modification of TTW balance by non-traditional effects}
\shortauthor{M. N. Crowe}
\affiliation{Department of Mathematics, University College London, London, WC1E 6BT, UK}
\date{}

\begin{document}

\maketitle

\begin{abstract}
The meridional component of the earth's rotation is often neglected in geophysical contexts. This is referred to as the `traditional approximation' and is justified by the typically small vertical velocity and aspect ratio of such problems. Ocean fronts are regions of strong horizontal buoyancy gradient and are associated with strong vertical transport of tracers and nutrients. Given these comparatively large vertical velocities, non-traditional rotation may play a role in governing frontal dynamics.

Here the effects of non-traditional rotation on a front in turbulent thermal wind balance are considered using an asymptotic approach. Solutions are presented for a general horizontal buoyancy profile and examined in the simple case of a straight front. Non-traditional effects are found to depend strongly on the direction of the front and may lead to the generation of jets and the modification of the frontal circulation and vertical transport.
\end{abstract}


\vspace{0.3cm}
\hspace{0.2cm}\rule{12.3cm}{0.4pt}

\section{Introduction}

The so-called `traditional approximation' \citep{ECKART,GERKEMAETAL,LUCAS_ETAL} describes the neglect of the meridional (North-South) component of the planetary rotation vector. This approximation is justified by a scaling argument and valid for flows in which the vertical length-scales are small compared to the horizontal length-scales and the vertical velocities are small. While the traditional approximation is generally accurate for oceanic and atmospheric flows, the effects of the neglected rotation component - referred to here as non-traditional effects - can still be important in some problems, particularly if the vertical velocities are large or the traditional rotation vector vanishes.

For flows with strong vertical velocities, non-traditional rotation can have a variety of effects such as introducing directional dependence in Ekman flows \citep{colemanetal,MCWILLIAMSHUCKLE} and tilting convective plumes in deep convection \citep{GARWOOD,sheremet_2004}. Near the equator, the traditional Coriolis parameter is small and non-traditional rotation dominates. This results in a different form of geostrophic balance \citep{verdiere_schopp_1994} in which horizontal density gradients are balanced by the meriodionally sheared velocity and can lead to the emergence of new phenomena such as the deep equatorial jets studied by \citet{hua_etal}.

Non-traditional effects also play an important role in the dynamics of internal waves \citep{gerkema_shrira_2005,GERKEMAETAL}, particularly in the case of near-inertial waves where they act as a singular perturbation, resulting in qualitatively different behaviour to the traditional system even when a scaling argument would suggest these effects are small. This perturbation corresponds to the existence of a range of trapped sub-inertial modes which vanish under the traditional approximation. Other effects include increasing the critical latitude at which internal waves can no longer propagate and modifying the reflection off a sloping bottom \citep{GERKEMA_2006}.

Ocean fronts are regions of strong horizontal buoyancy gradient and are common features in the upper ocean. These fronts typically occur on horizontal scales of around $1-10\,\textrm{km}$ and exist in a state close to turbulent thermal wind (TTW) balance - the three way balance between the Coriolis force, horizontal pressure gradients and the vertical mixing of momentum \citep{CRONINKESSLER,GULAETAL,MCWILLIAMSETAL,WENEGRAT}. Frontal systems are predominantly hydrostatic so vertical pressure gradients are set by the fluid density. An important dynamical feature of frontal systems is the secondary circulation \citep{MCWILLIAMS} which is associated with an enhanced vertical velocity and acts to exchange heat and nutrients \citep{GARRETTLODER,FERRARI} between the surface and the ocean interior. Due to this large vertical velocity, non-traditional effects may play a role in governing frontal dynamics.

\citet{CROWETAYLOR} considered a simple analytical model for a front in TTW balance. Vertical mixing was shown to generate a leading order cross-front flow which drives a circulation around the front and hence strong up/downwelling at the frontal edges. The circulation acts to restratify the front through the tilting of vertical buoyancy contours and the induced vertical stratification is maintained through an advection-diffusion balance. Over very long time-scales, the correlation between the cross-front flow and vertical stratification was shown to result in frontal spreading via shear dispersion. These predictions were tested in \citet{CROWETAYLOR3} and the model was extended to include the effects of surface wind stress and buoyancy flux in \citet{CROWETAYLOR20} and used to study the effects of vertical mixing on baroclinic instability in \citet{CROWETAYLOR2}.

Here, the effects of non-traditional rotation on a front in TTW balance are considered by including these effects as a perturbation from the TTW solution of \citet{CROWETAYLOR}. A small parameter representing the strength of the non-traditional rotation component is introduced and asymptotic solutions for the velocity fields and induced stratification are derived. The magnitude of the non-traditional correction terms is found to depend strongly on the angle of the front with fronts aligned in the East-West direction being most strongly affected by non-traditional rotation and fronts aligned in the North-South direction being unaffected.

An important feature of the solution is the generation of vertical vorticity by the horizontal component of the non-traditional Coriolis force. This vorticity appears as along-front jets and results in temporal evolution of the system over much faster timescales than the shear dispersion observed by \citet{CROWETAYLOR}. Additionally, it is found that non-traditional effects can modify the circulation around the front leading to enhanced vertical transport and regions of increased surface velocity convergence. This velocity convergence is frontogenetic \citep{HOSKINS,SHAKESPEARETAYLOR,MCWILLIAMS} - driving a sharpening of the horizontal buoyancy gradients - however it should be noted that the predicted sharpening is weak and non-traditional effects are unlikely to be a dominant mechanism for frontogenesis.

In \cref{sec:setup} the problem setup is described and the parameters and governing equations introduced. General asymptotic solutions are derived in \cref{sec:asymp} and summarised in \cref{sec:summ} with reference to the special case of a straight front. A specific example is illustrated in \cref{sec:example} and the features of the solution are shown and discussed. Finally in \cref{sec:diss} the results are discussed with reference to typical ocean parameters and areas for future work.

\section{Setup}
\label{sec:setup}

Consider a horizontally infinite layer of fluid between two rigid, horizontal boundaries with Cartesian coordinates $(x,y,z)$. Here $x$ describes the East-West direction, $y$ describes the North-South direction and $z$ is the vertical coordinate representing depth.The system is taken to be rotating with a constant angular velocity about the $y$ and $z$ axes. Evolution is governed by the incompressible Boussinesq equations where density changes are represented by a single scalar, buoyancy, with a single scalar equation describing its evolution. The governing equations can now be written \citep{CROWETAYLOR,CHARNEY} as
\begin{subequations}
\begin{alignat}{3}
\Dder{\textbf{u}}{t} + \textbf{f}\times\textbf{u} =& -\nabla p + b\hat{\textbf{z}}+ \nu \nabla^2 \textbf{u},\\
\nabla\cdot\textbf{u} = & \,0,\\
\label{eq:buoy_gov}
\Dder{b}{t} = & \,\kappa\nabla^2 b,
\end{alignat}
\end{subequations}
for
\begin{equation}
\textbf{f} = \begin{pmatrix} 0\\\tw{f}\\f\end{pmatrix},\quad\quad \hat{\textbf{z}} = \begin{pmatrix} 0\\0\\1\end{pmatrix},
\end{equation}
where $f$ and $\tw{f}$ describe the vertical and meridional components of rotation respectively. Due to the typically small horizontal scales of ocean fronts, the beta effect is not considered and $f$ and $\tw{f}$ are taken to be constant. Using typical horizontal lengthscale, $(x,y)\sim L$, typical buoyancy scale, $b\sim B$, inertial timescale, $t \sim 1/f$, and layer depth, $H$, it is convenient to nondimensionalise $(u,v)$ by $U = BH/(fL)$, $w$ by $BH^2/(fL^2)$ and $p$ by $BH$. The system is now described by five nondimensional parameters; the Rossby number, $\Ro = U/(fL)$, the Ekman number, $\E = \nu/(fH^2)$, the Prandtl number, $\Pr = \nu/\kappa$, the aspect ratio, $\epsilon = H/L$ and the ratio $\tw{f}/f$. It should be noted that $(f,\tw{f}) = 2\Omega(\sin\theta,\cos\theta)$ where $\Omega$ is the rotation rate of the Earth and $\theta$ is the latitude. Therefore
\begin{equation}
\frac{\tw{f}}{f} = \frac{1}{\tan\theta},
\end{equation}
so non-traditional effects will be amplified near the equator where $\theta$ is small. The ratio $\tw{f}/f$ only appears multiplied by $\epsilon$ so a non-traditional parameter
\begin{equation}
\delta = \frac{\epsilon \tw{f}}{f},
\end{equation}
is introduced for brevity. The governing equations can now be written as
\begin{subequations}
\label{eq:TTW}
\begin{alignat}{3}
\label{eq:TTW_a}
\pder{u}{t}+\Ro\left[\ader{u}{}\right]+\delta w - v = & -\pder{p}{x}+\E\pder{^2u}{z^2},\\
\label{eq:TTW_b}
\pder{v}{t}+\Ro\left[\ader{v}{}\right]\hspace{24pt}+ u = & -\pder{p}{y}+\E\pder{^2v}{z^2},\\
\label{eq:TTW_c}
\pder{b}{t}+\Ro\left[\ader{b}{}\right]\hspace{42pt}=&\hspace{35pt}\frac{\E}{\Pr}\pder{^2b}{z^2},\\
\label{eq:TTW_d}
-\delta u =& -\pder{p}{z}+b,\\
\label{eq:TTW_e}
\pder{u}{x}+\pder{v}{y}+\pder{w}{z} = &\,0,
\end{alignat}
\end{subequations}
where all terms scaled by $\epsilon^2$ have been neglected. Therefore, the vertical momentum equation reduces to quasi-hydrostatic balance and any horizontal mixing terms vanish. Top and bottom boundaries are placed at $z = \pm 1/2$ where no-stress conditions are imposed on the horizontal velocity, no-flow conditions on the vertical velocity and no-flux conditions on the buoyancy. These conditions are taken for simplicity and may be replaced by a wind stress or heat flux condition as considered by \citet{CROWETAYLOR20}.

In the following analysis the depth-dependent and depth-independent parts of fields are often considered separately so it is convenient to define the depth-average
\begin{equation}
\da{*} = \int_{-1/2}^{1/2} * \dint z,
\end{equation}
and denote the deviation from this depth average by $*' = * - \da{*}$. Additionally, the horizontal gradient vector is denoted by
\begin{equation}
\nabla_H = \left(\pder{}{x},\pder{}{y},0\right).
\end{equation}

\begin{figure}
	\centering
	\includegraphics[width=4in]{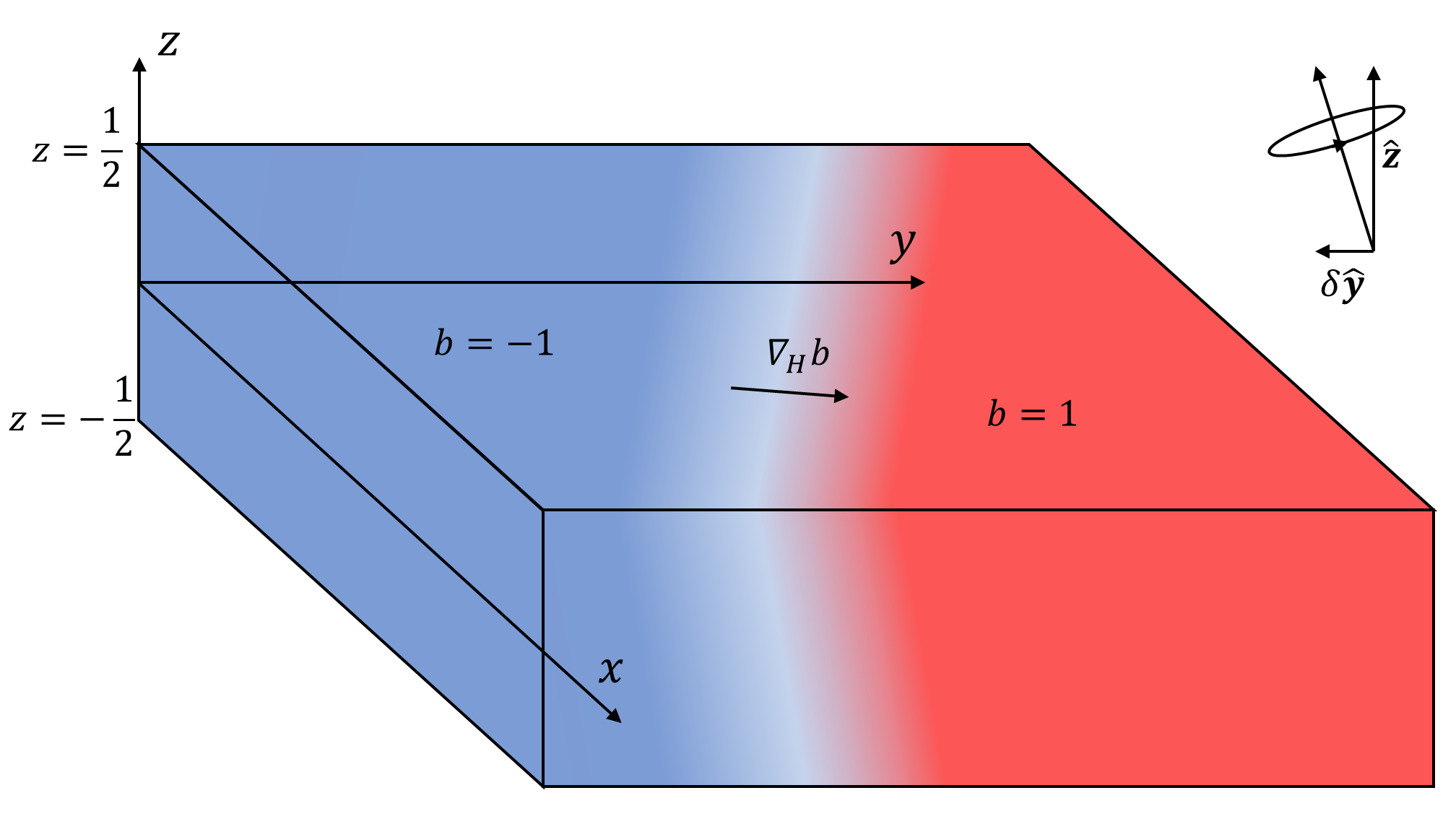}
	\caption{Typical non-dimensional frontal geometry showing a front with horizontal buoyancy gradient, $\nabla_H b$, and buoyancy of $b = -1$ (resp. $b = 1$) on the low (resp. high) buoyancy side of the front. Top and bottom boundary conditions are applied at $z = \pm 1/2$. In this non-dimensional setup, the system is rotating with angular velocity $\delta \hat{\textbf{y}}+\hat{\textbf{z}}$.}
    \label{fig:setup}
\end{figure}

An ocean front is represented here as an isolated region of non-zero horizontal buoyancy gradient, $\nabla_H b$, with $b = -1$ on the low buoyancy side and $b = 1$ on the high buoyancy side. The cross-front direction is defined to be the direction aligned with $\nabla_H b$ and the along-front direction to be aligned with $\hat{\textbf{z}}\times\nabla_H b$. Typically, variations in the along-front direction occur over larger scales than cross-front variations and hence examples of fronts with no along-front variation are used to illustrate these results. A typical frontal setup is shown in \cref{fig:setup}.

If the system is independent of $y$ - corresponding to a front aligned in the North-South direction - the non-traditional terms can be removed from \cref{eq:TTW} by replacing $p$ by $p+p_\delta$ where $p_\delta$ is defined using
\begin{equation}
\pder{p_\delta}{x} = -\delta w \quad \textrm{and} \quad \pder{p_\delta}{z} = \delta u.
\end{equation}
This definition is consistent as it can be easily shown to satisfy mass conservation. The resulting system is equivalent to setting $\delta = 0$ and hence non-traditional effects have no effect beyond the addition of an extra term in the pressure field.

\section{Asymptotic expansion}
\label{sec:asymp}

To proceed, the parameters $\delta$ and $\Ro$ are assumed small with $\delta \gg \Ro$. Taking $\Ro \sim \delta^2$, quantities may be expanded using an asymptotic expansion in $\delta$ by writing
\begin{equation}
\varphi = \varphi_0 + \delta\,\varphi_1 + \delta^2 \varphi_2 + \dots,
\end{equation}
for some field $\varphi$. Substituting expansions of this form into \cref{eq:TTW} gives a system of equations for each power of $\delta$. Typically, Ekman numbers lie in the range of $\E \sim 0.01 - 1$ \citep{CROWETAYLOR}. However, it should be noted that even for $\E \ll 1$, fields may be significantly modified within the top and bottom Ekman layers (of depth $O(\sqrt{\E})$) so $E$ is taken to be an $O(1)$ quantity throughout. Mathematically, this may be seen as retaining the highest vertical derivatives in order to enforce the top and bottom boundary conditions.

Before proceeding with the analysis it is worth discussing the time derivative terms in \cref{eq:TTW}. Unlike the TTW solutions of \citet{CROWETAYLOR,CROWETAYLOR3}, steady solutions to order $O(\Ro)$ do not exist; this unsteadiness results from the generation of depth-averaged vorticity by non-traditional effects.

\subsection{Generation of vorticity by non-traditional effects}
\label{sec:vor_gen}

Neglecting terms of order $O(\delta^2)$ from \cref{eq:TTW} and depth-averaging \cref{eq:TTW_a,eq:TTW_b,eq:TTW_e} gives
\begin{subequations}
\label{eq:TTW_vor}
\begin{alignat}{3}
\pder{\da{u}}{t}+\delta \da{w} - \da{v} = & -\pder{\da{p}}{x},\\
\pder{\da{v}}{t}+ \da{u} = & -\pder{\da{p}}{y},\\
\pder{\da{u}}{x}+\pder{\da{v}}{y} = &\,0,
\end{alignat}
\end{subequations}
which may be combined to give
\begin{equation}
\label{eq:vor_gen}
\pder{}{t}\left(\pder{\da{v}}{x}-\pder{\da{u}}{y}\right) = \delta \pder{\da{w}}{y}.
\end{equation}
\cref{eq:vor_gen} states that the non-traditional component of the Coriolis force acts to generate vorticity over long times, $t  \sim O(1/\delta)$. This suggests the inclusion of a second timescale, $T = \delta t$, corresponding to this vorticity generation. Using a multiple scales approach the time derivative may be expanded as
\begin{equation}
\pder{}{t} \to \pder{}{t}+\delta\pder{}{T},
\end{equation}
where now the $\pderline{}{t}$ term corresponds to transient inertial oscillations resulting from an unbalanced initial condition. From \citet{CROWETAYLOR} a longer timescale on the order of $t \sim O(\delta^4)$ is also expected to be important. This slow scale corresponds to shear dispersive spreading of the front and will be discussed in \cref{sec:high_ord_eff}.

From now on transient oscillations are neglected by setting the fast time derivative, $\pderline{}{t}$, to zero. Therefore the system is assumed to be balanced over the inertial timescale $t$ and only the slow evolution is considered.

\subsection{The $O(1)$ solution}

At leading order in $\delta$ \cref{eq:TTW} gives
\begin{subequations}
\label{eq:TTW_0}
\begin{alignat}{3}
- v_0 = & -\pder{p_0}{x}+\E\pder{^2u_0}{z^2},\\
u_0 = & -\pder{p_0}{y}+\E\pder{^2v_0}{z^2},\\
0 =&\hspace{35pt}\frac{\E}{\Pr}\pder{^2b_0}{z^2},\\
0 =& -\pder{p_0}{z}+b_0,\\
\pder{u_0}{x}+\pder{v_0}{y}+\pder{w_0}{z} = &\,0,
\end{alignat}
\end{subequations}
corresponding to the leading order (in $\Ro$) TTW system of \citet{CROWETAYLOR}. The leading order buoyancy equation may now be solved for
\begin{equation}
b_0 = b_0(x,y,T),
\end{equation}
hence the layer is vertically well mixed to leading order in $\delta$. The leading order pressure may now be solved as
\begin{equation}
p_0 = \da{p}_0+z\,b_0,
\end{equation}
where $\da{p}_0$ balances the depth-averaged component of velocity through geostrophic balance. This depth-averaged flow may be represented as a streamfunction by
\begin{equation}
\da{u}_0 = -\pder{\psi_0}{y},\quad \da{v}_0 = \pder{\psi_0}{x},
\end{equation}
where $\psi_0 = \da{p}_0$ so the depth-averaged pressure acts as a streamfunction for this horizontal flow. The depth-dependent velocity fields, $(u_0',v_0',w_0)$, may be calculated (see \citet{CROWETAYLOR}) by solving a fourth order linear system to obtain solution
\begin{subequations}
\label{eq:TTW_sol_0}
\begin{alignat}{3}
\label{eq:TTW0_u}
u_0' = & -\sqrt{\E}\left[K''(\z)\pder{b_0}{x}-K(\z)\pder{b_0}{y}\right],\\
v_0' = & -\sqrt{\E}\left[K(\z)\pder{b_0}{x}+K''(\z)\pder{b_0}{y}\right],\\
\label{eq:TTW0_w}
w_0 =& \,\,\E\, K'(\z) \nabla_H^2 b_0,
\end{alignat}
\end{subequations}
where $\z = z/\sqrt{\E}$ and $K(\z)$ is an $\E$ dependent vertical structure function satisfying
\begin{equation}
\begin{cases}
K^{(4)}(\z)+K(\z)+\z = 0 & \textrm{for}\hspace{7pt}  \z \in[-\z_0,\z_0],\\
K'(\z) = 0 & \textrm{at}\quad  \z = \pm \z_0,\\
K'''(\z) = 0 & \textrm{at}\quad  \z = \pm \z_0,
\end{cases}
\end{equation}
where $\z_0 = 1/(2\sqrt{\E})$ is the value of $|\z|$ on the top and bottom surfaces. Note that primes ($'$) on $K$ are taken to mean derivatives with respect to $\z$ rather than deviations from a vertical average as used elsewhere. The full solution for $K(\z)$ is given by $K_0(\z)$ in Appendix A of \citet{CROWETAYLOR}. For $\E \ll 1$ it can be shown that $K(\z)\sim-\z$ and hence thermal wind balance holds outside of thin boundary layers of width $O(\sqrt{\E})$ near the top and bottom boundaries.

\subsection{The $O(\delta)$ solution}

At order $O(\delta)$ \cref{eq:TTW} gives
\begin{subequations}
\label{eq:TTW_1}
\begin{alignat}{3}
\label{eq:TTW1_a}
\pder{u_0}{T}+w_0 - v_1 = & -\pder{p_1}{x}+\E\pder{^2u_1}{z^2},\\
\label{eq:TTW1_b}
\pder{v_0}{T}\hspace{24pt}+ u_1 = & -\pder{p_1}{y}+\E\pder{^2v_1}{z^2},\\
\label{eq:TTW1_c}
\pder{b_0}{T}=&\quad\frac{\E}{\Pr}\pder{^2b_1}{z^2},\\
\label{eq:TTW1_d}
- u_0 =& -\pder{p_1}{z}+b_1,\\
\label{eq:TTW1_e}
\pder{u_1}{x}+\pder{v_1}{y}+\pder{w_1}{z} = &\,0.
\end{alignat}
\end{subequations}
It can be shown that the only solutions satisfying \cref{eq:TTW1_c} along with no flux boundary conditions are
\begin{equation}
\pder{b_0}{T} = 0 \quad \textrm{and} \quad b_1 = b_1(x,y,T).
\end{equation}
Therefore $b_0$ does not change over the timescale $t = O(1/\delta)$ and the buoyancy is also depth-independent to $O(\delta)$. The pressure may now be calculated using \cref{eq:TTW1_d,eq:TTW0_u} as
\begin{equation}
\label{eq:p_1}
p_1 = \da{p}_1+(b_1 + \da{u}_0) z -\E\!\left[\left(\!K'(\z)-\frac{K(\z_0)}{\z_0}\!\right)\!\pder{b_0}{x} \!+\!\left(\!K'''(\z)-\frac{K''(\z_0)}{\z_0}\!+\!\frac{\z^2}{2}\!-\!\frac{\z_0^2}{6}\!\right)\!\pder{b_0}{y}\right]\!\!,
\end{equation}
where the final term arises from the integral of $u_0'$ and has been set to be depth-independent.

\subsubsection{The depth-averaged system}

From \cref{eq:TTW1_a,eq:TTW1_b,eq:TTW1_e}, the depth-averaged velocity and pressure satisfy
\begin{subequations}
\label{eq:TTW1_vor}
\begin{alignat}{3}
\label{eq:TTW1_vor_a}
\pder{\da{u}_0}{T}+ \da{w}_0 - \da{v}_1 = & -\pder{\da{p}_1}{x},\\
\label{eq:TTW1_vor_b}
\pder{\da{v}_0}{T}+ \da{u}_1 = & -\pder{\da{p}_1}{y},\\
\label{eq:TTW1_vor_c}
\pder{\da{u}_1}{x}+\pder{\da{v}_1}{y} = &\,0,
\end{alignat}
\end{subequations}
which may be combined to give
\begin{equation}
\pder{}{T}\left(\pder{\da{v}_0}{x}-\pder{\da{u}_0}{y}\right) = \pder{\da{w}_0}{y} \quad \implies \quad \pder{}{T}\nabla^2 \psi_0 = \pder{\da{w}_0}{y},
\end{equation}
which describes the generation of depth-averaged vorticity. Substituting for $\da{w}_0$ gives that
\begin{equation}
\pder{\psi_0}{T} = 2 \sqrt{\E^{3}}\, K(\z_0) \pder{b_0}{y} \quad \implies \quad \psi_0 = \Psi_0 + 2\sqrt{\E^{3}}\, K(\z_0) \pder{b_0}{y} T,
\end{equation}
where $\Psi_0 = \Psi_0(x,y)$ is the value of $\psi_0$ at $T = 0$. The depth-averaged geostrophic flow can now be determined from $\psi_0$. From \cref{eq:TTW1_vor_c} the $O(\delta)$ depth-averaged flow may now be written as
\begin{equation}
\da{u}_1 = -\pder{\psi_1}{y},\quad \da{v}_1 = \pder{\psi_1}{x},
\end{equation}
where, by \cref{eq:TTW1_vor_a,eq:TTW1_vor_b}, $\psi_1$ is related to $\da{p}_1$ through
\begin{equation}
\da{p}_1 = \psi_1 - 2\sqrt{\E^{3}}\,K(\z_0) \pder{b_0}{x}.
\end{equation}
To determine the evolution of $\psi_1$ it is necessary to consider the $O(\delta^2)$ system.

\subsubsection{The depth-dependent system}

The depth-dependent quantities may now be considered by subtracting the depth-averaged horizontal momentum equations in \cref{eq:TTW1_vor} from \cref{eq:TTW1_a,eq:TTW1_b} to obtain
\begin{subequations}
\label{eq:TTW1_step1}
\begin{alignat}{3}
w_0' - v_1' = & -\pder{p_1'}{x}+\E\pder{^2u_1'}{z^2},\\
u_1' = & -\pder{p_1'}{y}+\E\pder{^2v_1'}{z^2},
\end{alignat}
\end{subequations}
where the time derivative terms vanish as $(u_0',v_0')$ does not depend on $T$. Substituting for $w_0'$ using \cref{eq:TTW0_w} and $p_1'$ using \cref{eq:p_1}, this system may be solved (see \cref{sec:app_sol1}) for solution
\begin{subequations}
\begin{alignat}{3}
u_1' = & -\sqrt{\E}\left[K''(\z)\pder{}{x}-K(\z)\pder{}{y}\!\right]\!\left(b_1-\pder{\psi_0}{y}\right)+\E\pder{}{y}\left[A(\z)\pder{b_0}{x}-B(\z)\pder{b_0}{y}\right]\!,\\
v_1' = & -\sqrt{\E}\left[K(\z)\pder{}{x}+K''(\z)\pder{}{y}\!\right]\!\left(b_1-\pder{\psi_0}{y}\right)+\E\pder{}{y}\left[B(\z)\pder{b_0}{x}+A(\z)\pder{b_0}{y}\right]\!.
\end{alignat}
\end{subequations}
Finally, $w_1$ may be calculated  using \cref{eq:TTW1_e} as
\begin{equation}
w_1 = \E\, K'(\z) \nabla_H^2 \left(b_1 - \pder{\psi_0}{y} \right) - \sqrt{\E^3} \, C(\z) \nabla_H^2 \pder{b_0}{y},
\end{equation}
where $C(\z)$ is the integral of $A(\z)$. The functions $A$, $B$ and $C$ are complicated functions of $\z$, $K(\z)$ and $\z_0$ and are given in \cref{sec:app1}.

\subsection{The $O(\delta^2)$ solution}

In \citet{CROWETAYLOR} it was shown that an $O(\Ro)$ stratification is induced and maintained by an advection-diffusion balance in the buoyancy equation. Here this effect is expected to appear at orders $O(\delta^2) = O(\Ro)$ and $O(\delta^3)$ and the $O(\delta^2)$ system is considered first.

\subsubsection{The buoyancy field}

Since it has been assumed that $\Ro = O(\delta^2)$, it is convenient to define $\Ro = \mathcal{R}\, \delta^2$ where $\mathcal{R}$ is an $O(1)$ number. The $O(\delta^2)$ buoyancy equation is
\begin{equation}
\label{eq:buoy_2}
\pder{b_1}{T} + \mathcal{R} \left(u_0\pder{b_0}{x}+v_0\pder{b_0}{y}\right) = \frac{\E}{\Pr} \pder{^2 b_2}{z^2},
\end{equation}
and noting that $b_1$ is depth-independent, \cref{eq:buoy_2} may be depth-averaged to obtain
\begin{equation}
\label{eq:buoy_2_da}
\pder{b_1}{T} + \mathcal{R} J(\psi_0,b_0) = 0,
\end{equation}
where $J(\phi,\varphi) = (\partial_x \phi)(\partial_y \varphi) - (\partial_y \phi)(\partial_x \varphi)$ is the Jacobian derivative. Substituting for $\psi_0$ gives
\begin{equation}
\label{eq:b1_sol}
b_1 = -\mathcal{R} \left[ J(\Psi_0,b_0)T + \sqrt{\E^3}\, K(\z_0)\, J\left(\pder{b_0}{y},b_0\right) T^2\right],
\end{equation}
assuming that $b_1 = 0$ at $T=0$.

Subtracting \cref{eq:buoy_2_da} from \cref{eq:buoy_2} gives
\begin{equation}
\mathcal{R}\left(u_0'\pder{b_0}{x}+v_0'\pder{b_0}{y}\right) = \frac{\E}{\Pr} \pder{^2 b_2}{z^2}.
\end{equation}
This equation was considered in \citet{CROWETAYLOR} and describes the restratification of the front by the TTW circulation. The solution is
\begin{equation}
\label{eq:b_sol_2}
b_2 = \da{b}_2(x,y,T) - \mathcal{R}\,\Pr\, \sqrt{\E}\, K(\z) |\nabla_H b_0|^2.
\end{equation}

\subsubsection{The streamfunction for the depth-averaged flow}

Depth-dependent velocity components of order higher than $O(\delta)$ are not required in the subsequent calculations. However, higher order components of $\psi$ are required to determine the higher order depth-averaged buoyancy terms and may be determined by considering the vertical vorticity.

The depth-averaged vertical vorticity equation may be derived by cross-differentiating \cref{eq:TTW_a} and \cref{eq:TTW_b} and depth-averaging to obtain
\begin{equation}
\label{eq:vort_eqn}
\delta \pder{\da{\eta}}{T} + \Ro\, \nabla_H \cdot \left[ \da{\textbf{u}_H \eta} - \da{\boldsymbol\omega_H w} \right] = \delta \pder{\da{w}}{y}.
\end{equation}
Here $\textbf{u}_H = (u,v,0)$ is the horizontal velocity, $\eta = \pderline{v}{x}-\pderline{u}{y}$ is the vertical vorticity and 
\begin{equation}
\boldsymbol\omega_H = \begin{pmatrix}\pder{w}{y}-\pder{v}{z}\\\pder{u}{z}-\pder{w}{x}\\0\end{pmatrix},
\end{equation}
is the horizontal vorticity. At $O(\delta^2)$ \cref{eq:vort_eqn} gives
\begin{equation}
\label{eq:psi_2}
\pder{\nabla_H^2\psi_1}{T}+\mathcal{R}\, J\left[\psi_0,\nabla_H^2 \psi_0\right] = \mathcal{R}\, \nabla_H \cdot \left[ -\da{\textbf{u}_{H0}' \eta_0'} + \da{\boldsymbol\omega_{H0} w_0} \right] + \pder{\da{w}_1}{y},
\end{equation}
where the flux terms can be expressed in terms of $b_0$ to give
\begin{equation}
\label{eq:psi_1_evol}
\pder{\nabla_H^2\psi_1}{T}+\mathcal{R}\, J\left[\psi_0,\nabla_H^2 \psi_0\right] = \mathcal{R}\, \nabla_H \cdot \left[ \mathsfbi{P}\cdot \nabla_H b_0 \, \nabla_H^2 b_0 \right] + \pder{\da{w}_1}{y},
\end{equation}
for
\begin{equation}
\mathsfbi{P} = \E \begin{pmatrix} 2\da{K'^2} & \da{K^2}-\da{K''^2} \\ \da{K''^2}-\da{K^2} & 2\da{K'^2} \end{pmatrix}.
\end{equation}
The flux term in \cref{eq:psi_1_evol} corresponds to both the generation of vorticity due to vortex stretching and the horizontal transport of vorticity due to a correlation between the vertically sheared profiles for the horizontal velocity and the vertical vorticity. Over timescales longer than $T$, these terms have been shown to generate along front jets \citep{CROWETAYLOR3} and play a role in baroclinic instability \citep{CROWETAYLOR2}. Vorticity is also generated by the non-traditional component of the Coriolis force through the $y$ variations in $\da{w}_1$, as discussed in \cref{sec:vor_gen}.

\cref{eq:psi_1_evol} may be solved for $\nabla_H^2\psi_1$ by a simple integration in $T$. However, solving for $\psi_1$ requires inverting the Laplacian operator so it is not possible to present a simple analytic solution. Solutions for \cref{eq:psi_1_evol} could be easily found numerically for given fields $b_0$, $\psi_0$ and $b_1$.

\subsection{The $O(\delta^3)$ solution}

Now the order $O(\delta^3)$ balance in considered to determine the stratification maintained by the $O(\delta)$ velocity component. The $O(\delta^3)$ vorticity equation will not be examined though it may be derived from \cref{eq:vort_eqn} similarly to \cref{eq:psi_1_evol}. The buoyancy equation is
\begin{equation}
\label{eq:buoy_3}
\pder{b_2}{T} + \mathcal{R} \left( u_1 \pder{b_0}{x}+v_1\pder{b_0}{y} + u_0\pder{b_1}{x}+v_0\pder{b_1}{y}\right) = \frac{\E}{\Pr} \pder{^2 b_3}{z^2},
\end{equation}
which may be depth-averaged to obtain
\begin{equation}
\label{eq:buoy_3_da}
\pder{\da{b}_2}{T} + \mathcal{R} \left[ J(\psi_1,b_0) + J(\psi_0,b_1) \right] = 0.
\end{equation}
This equation may be solved using the expression for $\psi_1$ if required. Since the depth-averaged buoyancy is known to the first two orders in $\delta$ and it is not possible to find a simple analytic expression for $\psi_1$, expressions for $\da{b}$ are not calculated explicitly at $O(\delta^2)$ or higher. Instead, the focus is on determining the vertical structure of $b$, denoted $b'$, to the lowest two orders. Since the lowest order term in $b'$ is $b'_2$ (see \cref{eq:b_sol_2}), the next order term, $b_3'$, must also be determined.

Subtracting \cref{eq:buoy_3_da} from \cref{eq:buoy_3} and noting that $\pderline{b_2'}{T} = 0$ gives the equation for the depth-dependent buoyancy
\begin{equation}
\mathcal{R} \left( u_1' \pder{b_0}{x} + v_1' \pder{b_0}{y} + u_0' \pder{b_1}{x} + v_0' \pder{b_1}{y} \right) = \frac{\E}{\Pr}\pder{^2 b_3}{z^2},
\end{equation}
with solution
\begin{multline}
\label{eq:b3_sol}
b_3 = \da{b}_3(x,y,T) + \mathcal{R}\,\Pr \left[ \E \left( \frac{D_1(\z)}{2} \pder{}{y} |\nabla_H b_0|^2 + D_2(\z)\, J\left[ \pder{b_0}{y},b_0\right] \right) \right. + \\ \left.\sqrt{\E} \left( K(\z) \left( \nabla_H \pder{\psi_0}{y} \cdot \nabla_H b_0 - 2 \nabla_H b_0 \cdot \nabla_H b_1 \right) -\left( K''(\z)+\frac{\z^3}{6}-\frac{\z\z_0^2}{2} \right) J\left[\pder{\psi_0}{y},b_0\right]\right)\right],
\end{multline}
where the vertical structure functions $D_1(\z)$ and $D_2(\z)$ are given in \cref{sec:app1}. The term $\da{b}_4$ can be determined by depth-averaging the $O(\delta^5)$ buoyancy equation, as noted above, this calculation is not done here.

\subsection{Higher order terms and shear dispersive spreading}
\label{sec:high_ord_eff}

The asymptotic approach may be continued as above to $O(\delta^4)$ and higher. However, from \citet{CROWETAYLOR,CROWETAYLOR3}, slow frontal spreading is expected due to a buoyancy flux resulting from the correlation between the leading order velocity and the $O(\delta^2)$ stratification, $\da{\textbf{u}_{H0}' b_2'}$. This spreading is due to shear dispersion and was found to appear in the equations at $O(\Ro^2) = O(\delta^4)$ and occur over a timescale of $t = O(1/\Ro^2) = O(1/\delta^4)$. Similarly, the flux terms $\da{\textbf{u}_{H1}' b_2'}$ and $\da{\textbf{u}_{H0}' b_3'}$ resulting from non-traditional effects might be expected to drive some buoyancy change at $O(\delta^5)$. Therefore, new timescales are introduced to examine the effects of this shear dispersion.

The timescale $T = \delta t$ was shown in \cref{sec:vor_gen} to be the timescale over which an $O(1)$ amount of depth-averaged vorticity is generated by non-traditional effects. Over timescales longer than $T$, many of the terms in $\da{b}$ and $\psi$ demonstrate secular growth and as such it is necessary to introduce additional slow timescales corresponding to this slow frontal spreading. This is done by equating the size of the time derivative of the leading order buoyancy, $b_0$, with the shear dispersion terms
\begin{equation}
\pder{b_0}{t} \sim \delta^4 \mathcal{R}\, \nabla_H \cdot \left[ \da{\textbf{u}_{H0}' b_2'} \right] \quad \textrm{and} \quad \pder{b_0}{t} \sim \delta^5 \mathcal{R}\, \nabla_H \cdot \left[ \da{\textbf{u}_{H1}' b_2'} + \da{\textbf{u}_{H0}' b_3'} \right],
\end{equation}
to get two timescales, $T_4 = \delta^4 t$ and $T_5 = \delta^5 t$, and letting $b_0$ depend on $T_4$ and $T_5$. Here $T_4$ corresponds to the slow spreading timescale from \citet{CROWETAYLOR} while $T_5$ corresponds to a longer timescale on which the evolution of depth-averaged buoyancy occurs due to non-traditional effects. 
Determining a closed system in full generality requires knowing how $\psi$ evolves over the slow scales $T_4$ and $T_5$ which requires examining high order equations for the depth-averaged vorticity \citep{CROWETAYLOR2,CROWETAYLOR3}. Instead, the simplifying assumption of a straight front is made. Under this assumption, the $\psi$ dependent terms 
vanish and equations purely in terms of $b_0$ are recovered as 
\begin{equation}
\label{eq:spreading_4}
\pder{b_0}{T_4} = \mathcal{R}^2\,\Pr\, \nabla_H \cdot \left[ \mathsfbi{Q} \cdot \nabla_H b_0 |\nabla_H b_0|^2\right],
\end{equation}
and
\begin{equation}
\label{eq:spreading_5}
\pder{b_0}{T_5} = \mathcal{R}^2\,\Pr\, \nabla_H \cdot \left[ \mathsfbi{R}_1 \cdot \pder{\nabla_H b_0}{y} |\nabla_H b_0|^2 + \mathsfbi{R}_2 \cdot \nabla_H b_0 \pder{}{y}|\nabla_H b_0|^2\right],
\end{equation}
where
\begin{equation}
\label{eq:Q_def}
\mathsfbi{Q}(E) = \E \begin{pmatrix} \da{K'^2} & \da{K^2} \\ -\da{K^2} & \da{K'^2} \end{pmatrix},
\end{equation}
and
\begin{equation}
\label{eq:R_def}
\mathsfbi{R}_1(E) = \E\sqrt{\E} \begin{pmatrix} \da{AK} & -\da{BK} \\ \da{BK} & \da{AK} \end{pmatrix}, \quad \mathsfbi{R}_2(E) = \frac{\E\sqrt{\E}}{2} \begin{pmatrix} \da{AK} & -\da{D_1K} \\ \da{D_1K} & \da{AK} \end{pmatrix}.
\end{equation}
\cref{eq:spreading_4} is identical to the result derived in \citet{CROWETAYLOR} and describes the spreading of a front due to a horizontal buoyancy flux resulting from the correlation between the induced stratification and the cross-front flow. \cref{eq:spreading_5} similarly describes a horizontal buoyancy flux, with terms arising from the non-traditional corrections to the stratification and cross-front flow.

It is worth noting that over long timescales the generation of significant background vorticity is expected, both by non-traditional effects as discussed in \cref{sec:vor_gen} and due to the correlation between along-front and cross-front velocity fields as shown in \cref{eq:psi_1_evol} and discussed in \citet{CROWETAYLOR3}. These correlation terms appear as a consequence of vertical mixing driving a cross-front flow and do not appear in the limit of $\E \to 0$. The generated vorticity manifests as along-front jets and can become large enough to significantly modify the absolute vorticity of the system resulting in a modification of the TTW velocity solution and hence a modified stratification and frontal spreading. Additionally, frontal systems are susceptible to baroclinic instability \citep{STONE1966,CROWETAYLOR2} which may lead to a breakdown of the straight front assumption.

\section{Summary of solution}
\label{sec:summ}

Here the solution of \cref{sec:asymp} is summarised and results are presented and discussed for a simple frontal geometry.

\subsection{The velocity fields}

Correct to $O(\delta)$, the velocity fields are given by
\begin{equation}
\textbf{u}_H = -\nabla\times\left[ (\psi_0+\delta\,\psi_1)\hat{\textbf{z}}\right]-\sqrt{\E}\, \mathsfbi{K}\cdot \nabla_H\! \left(\!b_0+\delta\,b_1 - \delta\pder{\psi_0}{y}\!\right) + \delta\,\E\, \mathsfbi{A}\cdot \nabla_H\pder{b_0}{y} + O(\delta^2),
\end{equation}
and
\begin{equation}
w = \E\, K'(\z) \nabla_H^2 \left(b_0+\delta\,b_1 - \delta\pder{\psi_0}{y}\right) - \delta \sqrt{\E^3} \, C(\z) \nabla_H^2 \pder{b_0}{y} +O(\delta^2),
\end{equation}
where
\begin{equation}
\mathsfbi{K}(\z) = \begin{pmatrix} K''(\z) & -K(\z) \\ K(\z) & K''(\z) \end{pmatrix} \quad \textrm{and} \quad \mathsfbi{A}(\z) = \begin{pmatrix} A(\z) & -B(\z) \\ B(\z) & A(\z) \end{pmatrix}.
\end{equation}
The depth-averaged velocity is described by a streamfunction where
\begin{equation}
\psi_0 = \Psi_0 + 2\sqrt{\E^{3}}\,\delta\,t\, K(\z_0) \pder{b_0}{y},
\end{equation}
for some initial streamfunction $\psi_0 = \Psi_0$ at $t = 0$. It should be noted that $t = O(1/\delta)$ so all terms here are leading order. The $O(\delta)$ streamfunction component, $\psi_1$, satisfies \cref{eq:psi_1_evol}.

The leading order flow can be split into components in the cross-front direction (described by the diagonal terms in $\mathsfbi{K}$) and along-front direction (describes by the off-diagonal terms in $\mathsfbi{K}$). However, the $O(\delta)$ terms are aligned relative to gradients of the north-south ($y$) derivatives of $b_0$ and $\psi_0$ which do not necessarily correspond to the direction of $\nabla_H b_0$.

Two special cases are $b_0 = b_0(x)$ and $b_0 = b_0(y)$. The case of $b_0 = b_0(x)$ describes a front with the along-front direction aligned with North-South. In this case all $y$ derivatives can be neglected and the non-traditional terms have no effect on the front as discussed in \cref{sec:setup}. Conversely, $b_0 = b_0(y)$ describes a front with the along-front direction aligned with East-West. In this case non-traditional effects are maximised and the gradients of $b_0$ are aligned with the gradients of $\pderline{b_0}{y}$ so the horizontal velocity terms driven by the non-traditional rotation can be easily split into cross-front and along-front components similarly to the leading order flow.

\subsection{The buoyancy field}

The buoyancy field can be split into depth-averaged and depth-dependent components. Correct to the lowest two orders in $\delta$ the solutions are
\begin{equation}
\label{eq:b_summ}
\da{b} = b_0+\delta\, b_1 +O(\delta^2) = b_0 -\Ro\,t\, J\left[\Psi_0 + \sqrt{\E^3}\,\delta\,t\, K(\z_0)\pder{b_0}{y},b_0\right] + O(\delta^2),
\end{equation}
where $\Ro\,t = O(\delta)$. The depth-dependent buoyancy is given by
\begin{multline}
\label{eq:bp_summ}
b' = \Ro\,\Pr\, \sqrt{\E} \left[ \left(- K(\z) +\delta\,\sqrt{\E} \frac{D_1(\z)}{2} \pder{}{y} \right) |\nabla_H b_0|^2 + \delta \left( \sqrt{\E} D_2(\z)\, J\left[ \pder{b_0}{y},b_0 \right] \right.\right. + \\ \left.\left.  K(\z) \nabla_H\left( \pder{\psi_0}{y}  - 2 b_1 \right)\cdot\nabla_H b_0 -\left( K''(\z)+\frac{\z^3}{6}-\frac{\z\z_0^2}{2} \right) J\left[\pder{\psi_0}{y},b_0\right]\right)\right] + O(\delta^4).
\end{multline}
Similarly to the velocity fields, if $b_0 = b_0(x)$ then the non-traditional rotation has no effect on the front and the solution reduces to the results of \citep{CROWETAYLOR}. From $b'$ the vertical buoyancy gradient, $N^2$, may be determined as
\begin{multline}
N^2 = \pder{b'}{z} = \Ro\,\Pr\left[ \left(- K'(\z) +\delta\,\sqrt{\E} \frac{D_1'(\z)}{2} \pder{}{y} \right) |\nabla_H b_0|^2 + \delta \left( \sqrt{\E} D_2'(\z)\, J\left[ \pder{b_0}{y},b_0 \right] \right.\right. + \\ \left.\left.  K'(\z) \nabla_H\left( \pder{\psi_0}{y}  - 2 b_1 \right)\cdot\nabla_H b_0 -\left( K'''(\z)+\frac{\z^2}{2}-\frac{\z_0^2}{2} \right) J\left[\pder{\psi_0}{y},b_0\right]\right)\right] + O(\delta^4).
\end{multline}
The horizontal buoyancy gradient may be similarly calculated using
\begin{equation}
M^2 = \nabla_H b = \nabla_H \da{b} + \nabla_H b',
\end{equation}
where the first term on the right-hand side is leading order and depth-independent while the second term is order $O(\Ro)$ and depth-dependent.

\subsection{Frontal spreading and shear dispersion}

Over very long times the front is expected to evolve through shear dispersion. \cref{eq:spreading_4,eq:spreading_5} may be combined to give
\begin{equation}
\label{eq:b_0_spread}
\pder{b_0}{t} = \Ro^2\,\Pr\, \nabla_H \cdot \left[ \left(\mathsfbi{Q} \cdot \nabla_H b_0+\delta\,\mathsfbi{R}_1 \cdot \nabla_H \pder{b_0}{y} + \delta\,\mathsfbi{R}_2 \cdot \nabla_H b_0\, \pder{}{y}\right)|\nabla_H b_0|^2\right],
\end{equation}
which is valid for a straight front provided the vorticity generated by non-traditional effects and vertical mixing is less than the background vorticity. Expressions for $\mathsfbi{Q}$, $\mathsfbi{R}_1$ and $\mathsfbi{R}_2$ are given in \cref{eq:Q_def,eq:R_def}. It should be noted that \cref{eq:b_0_spread} reduces to the results of \citet{CROWETAYLOR} for $b_0 = b_0(x)$, similarly to the results for velocity and buoyancy. If $b_0 = b_0(y)$ is an odd function of $y$, then solutions to \cref{eq:b_0_spread} will remain odd in $y$ for all time for the case of $\delta = 0$. However, for $\delta \neq 0$, the addition of an extra $y$ derivative in the non-traditional correction terms leads to an asymmetry and hence different evolution on each side of the front.

\section{A simple frontal geometry}
\label{sec:example}

To illustrate the results given in \cref{sec:summ}, solutions are plotted for the simple case of
\begin{equation}
\label{eq:b0_example}
b_0 = \tanh y.
\end{equation}
As noted in the previous section, this corresponds to a front with the along-front direction (here the $x$ direction) aligned with East-West so that non-traditional effects are maximised. From \cref{eq:b_summ} it can be seen that $b_1 = 0$ since the Jacobian terms vanish. Similarly, higher order depth-averaged buoyancy terms, such as $\da{b}_2$ and $\da{b}_3$, will evolve through advection by Jacobian terms so may also be set to zero. Therefore $b_0$ may be taken to describe the full depth-averaged buoyancy.

\subsection{Depth-independent jets}

\begin{figure}
	\centering
	\begin{subfigure}[b]{0.49\textwidth}
	\centering
	\includegraphics[trim={0 0 0 0},clip,width=\textwidth]{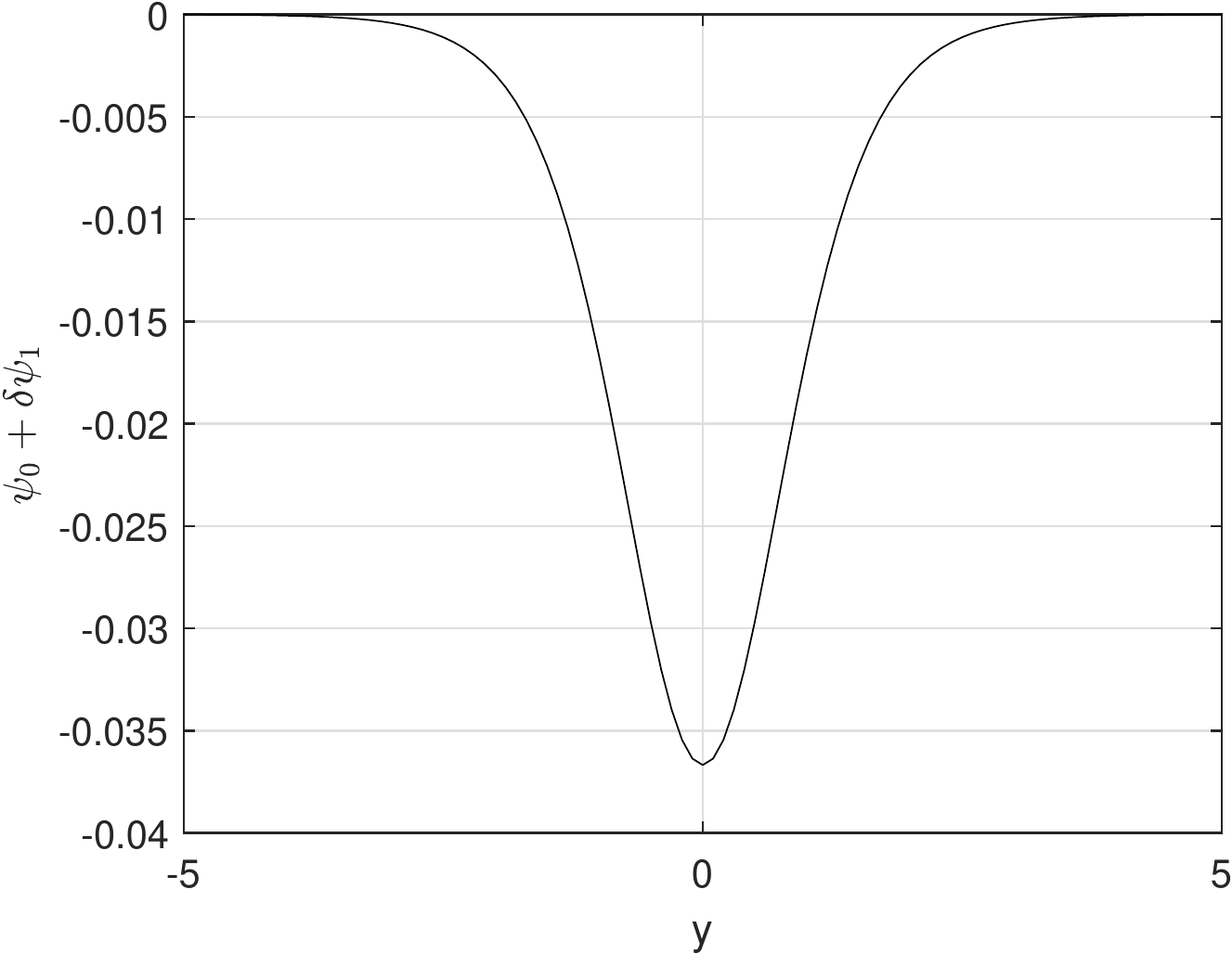}
	\caption{}
	\end{subfigure}
	\begin{subfigure}[b]{0.49\textwidth}
	\centering
	\includegraphics[trim={0 0 0 0},clip,width=\textwidth]{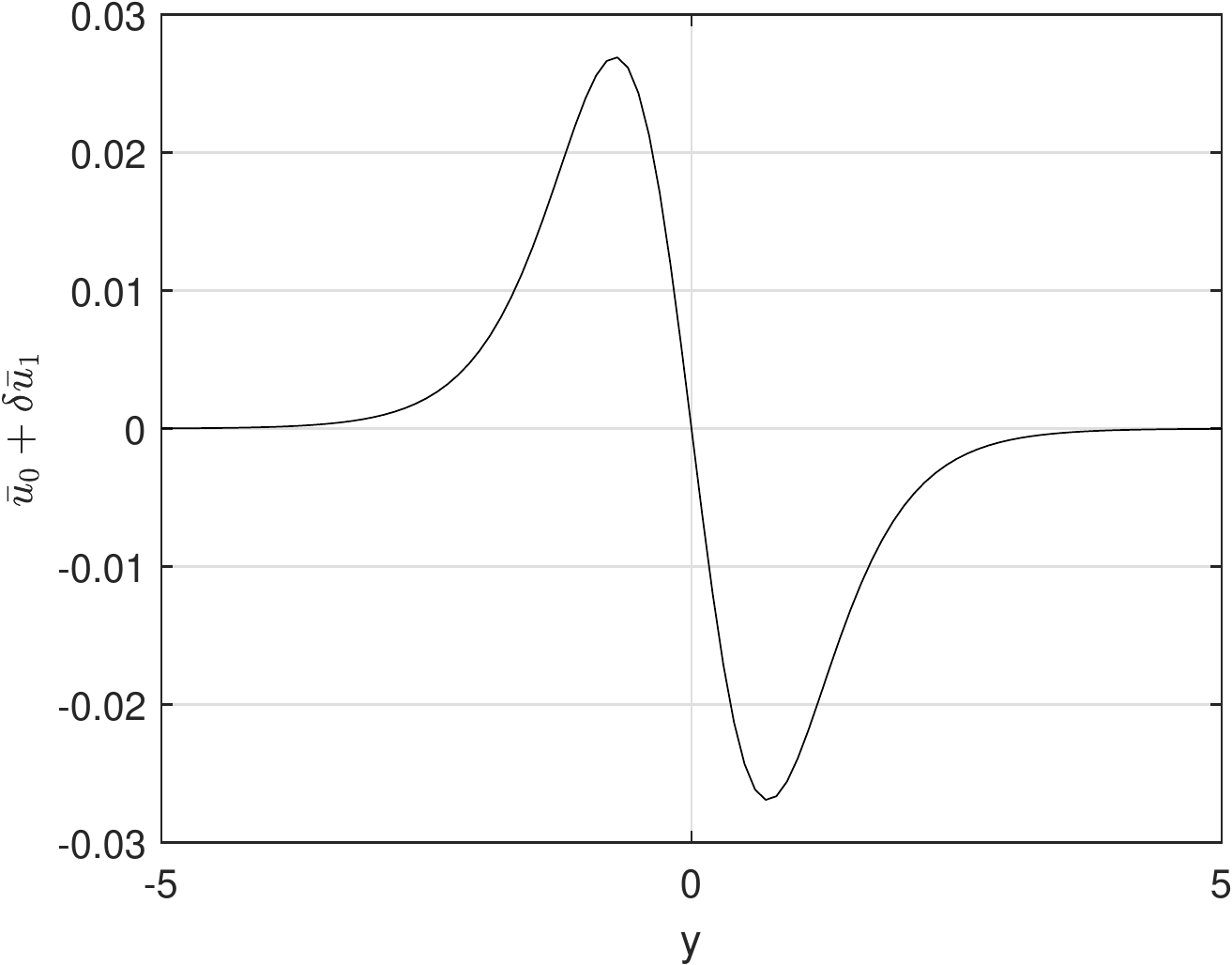}
	\caption{}
	\end{subfigure}
	\caption{The streamfunction (a) and velocity (b) of the along-front jets. Solutions are shown correct to $O(\delta)$ for $\E = 0.1$, $\delta = 0.2$, $\delta\,t = 1$ and $\mathcal{R} = 1$.}
    \label{fig:psi}
\end{figure}

Taking initial streamfunction of $\Psi_0 = 0$, the depth-averaged velocity is given by
\begin{equation}
\label{eq:psi0_example}
\psi_0 = 2\sqrt{\E^{3}}\, \delta\, t\, K(\z_0)\, \sech^2 y \quad \implies \quad \left(\da{u}_0,\da{v}_0\right) = 4\sqrt{\E^{3}}\, \delta\, t\, K(\z_0)\left(\sech^2 y \tanh y,\, 0\right),
\end{equation}
corresponding to two jets running in opposite directions along the edges of the front. As expected, motion is confined to the frontal region. Since the Jacobian terms vanish for $b_0 = b_0(y)$, \cref{eq:psi_1_evol} may be solved for $\psi_1$ as
\begin{equation}
\label{eq:psi_1_example}
\psi_1 = \mathcal{R}\,\E\,\delta\, t\, \da{K'^2}\left(\pder{b_0}{y}\right)^2 - 2\,\E^3\,(\delta\, t)^2\, [K(\z_0)]^2 \pder{^3b_0}{y^3}.
\end{equation}
The first term of $\psi_1$ in \cref{eq:psi_1_example} describes the vorticity generated by the correlation between the cross-front and along front TTW velocities \citep{CROWETAYLOR3} while the second term describes the generation of vorticity through the action of the non-traditional Coriolis force on the $O(\delta)$ vertical velocity. The streamfunction and along-front velocity of the depth-independent jets are shown in \cref{fig:psi} correct to $O(\delta)$ as a function of $y$ for $\E = 0.1$, $\delta = 0.2$, $\delta\,t = 1$ and $\mathcal{R} = 1$. These jets grow with time and are expected to become large for $T\gg 1$.

\subsection{Frontal circulation}

For an $x$ independent front, the cross-front velocity ($v$) and vertical velocity ($w$) satisfy the mass conservation equation
\begin{equation}
\pder{v'}{y} + \pder{w}{z} = 0, 
\end{equation}
and hence the circulation around the front in the $y-z$ plane can be represented by a circulation streamfunction, $\phi$, defined using
\begin{equation}
v' = \pder{\phi}{z}, \quad w = -\pder{\phi}{y}.
\end{equation}
Note that there is no depth-independent flow in the $y$ direction as $\psi = \psi(y)$ so $v = v'$ here. The circulation components, $\phi_0$ and $\phi_1$, are given by
\begin{equation}
\label{eq:phi_sol}
\phi_0 = -\E\, K'(\z) \pder{b_0}{y}, \quad \phi_1 = \sqrt{\E^3}\, C(\z) \pder{^2b_0}{y^2} + \E\, K'(\z)\pder{^2\psi_0}{y^2}.
\end{equation}
The two terms of $\phi_1$ in \cref{eq:phi_sol} each arise due to different components of the non-traditional Coriolis force. The horizontal component appears directly in the horizontal momentum balance resulting in the first term of \cref{eq:phi_sol} while the vertical component drives the system out of hydrostatic balance, modifying the pressure field and giving the second term.

\begin{figure}
	\centering
	\begin{subfigure}[b]{0.49\textwidth}
	\centering
	\includegraphics[trim={0 0 0 0},clip,width=\textwidth]{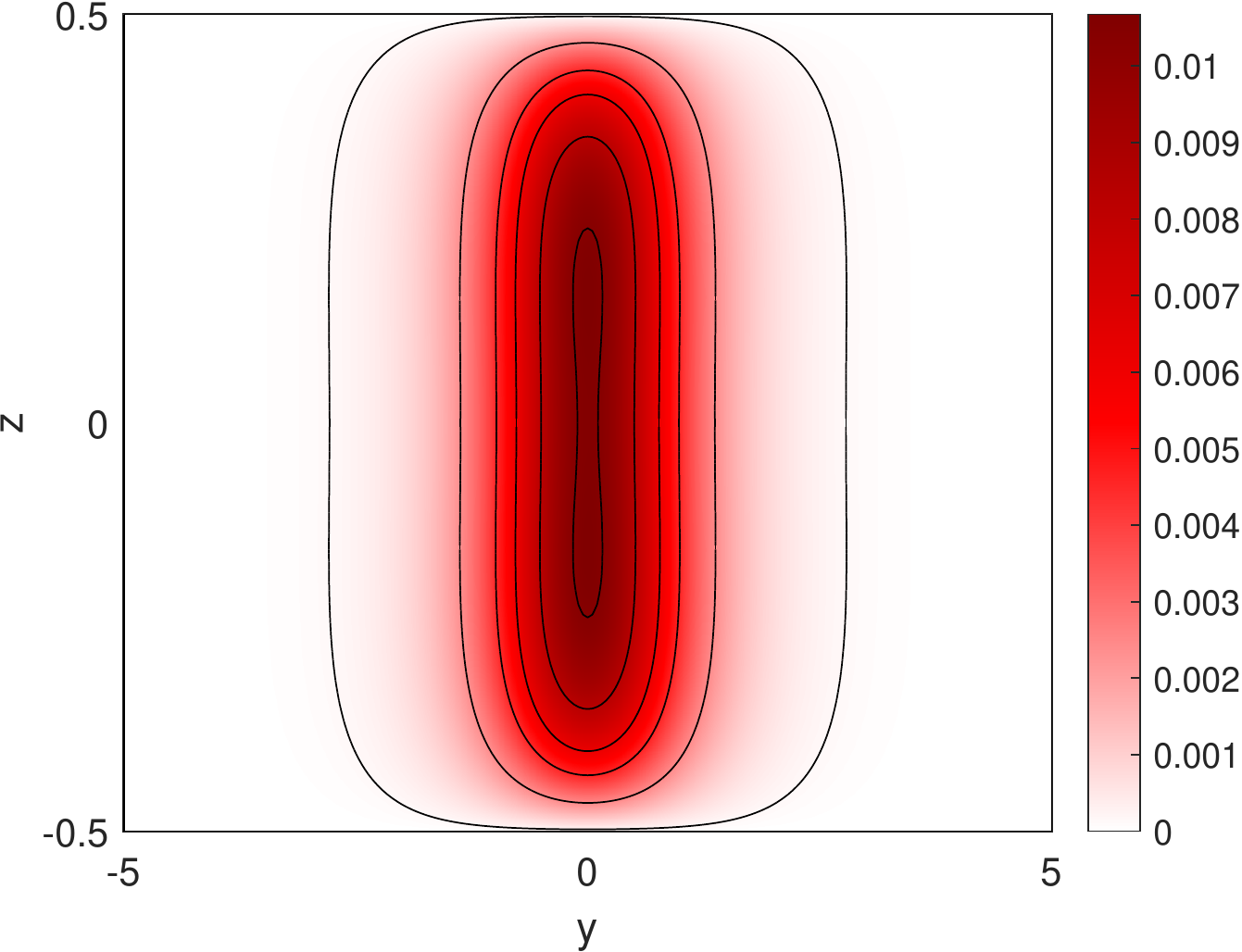}
	\caption{}
	\end{subfigure}
	\begin{subfigure}[b]{0.49\textwidth}
	\centering
	\includegraphics[trim={0 0 0 0},clip,width=\textwidth]{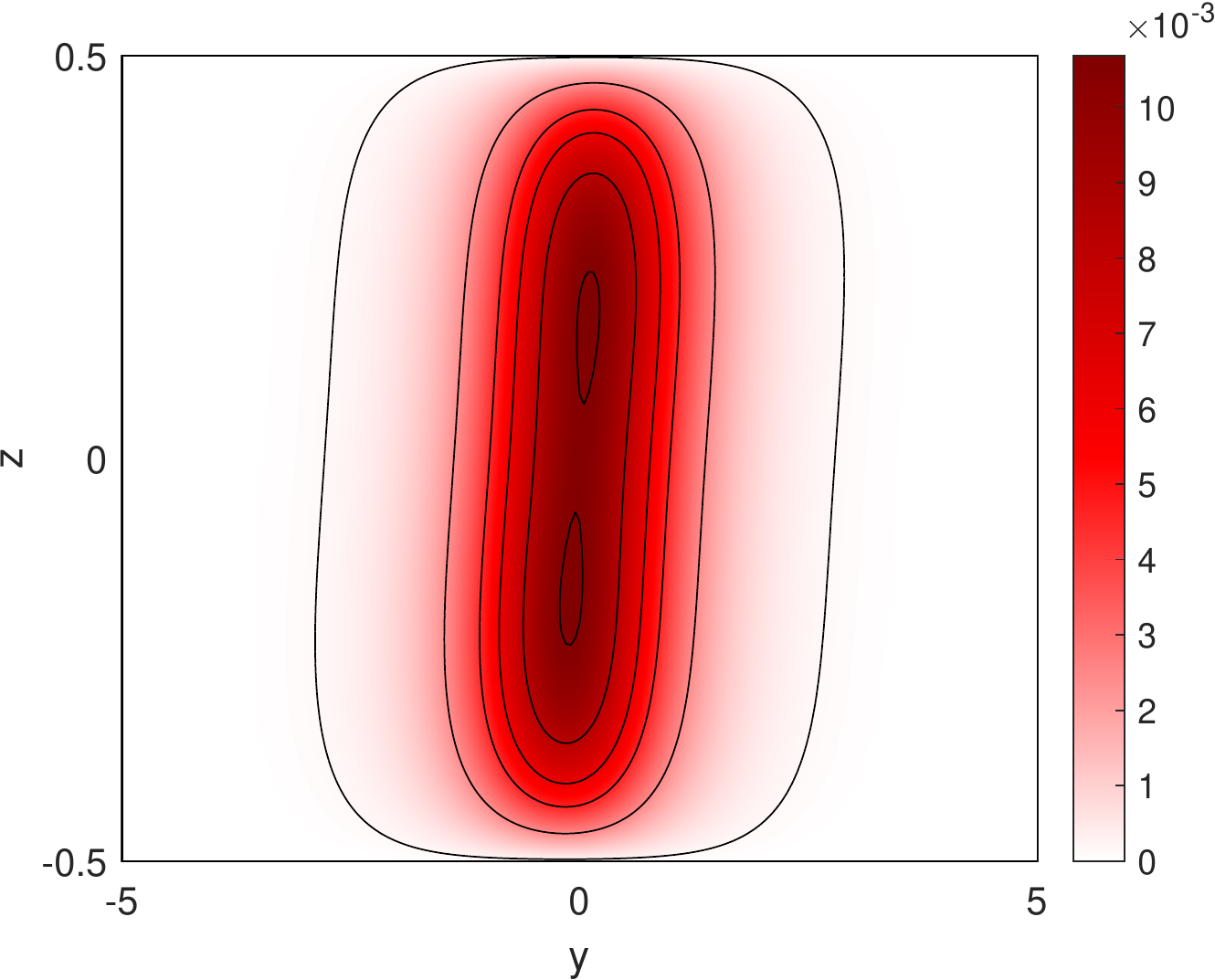}
	\caption{}
	\end{subfigure}
	\newline
	\begin{subfigure}[b]{0.49\textwidth}
	\centering
	\includegraphics[trim={0 0 0 0},clip,width=\textwidth]{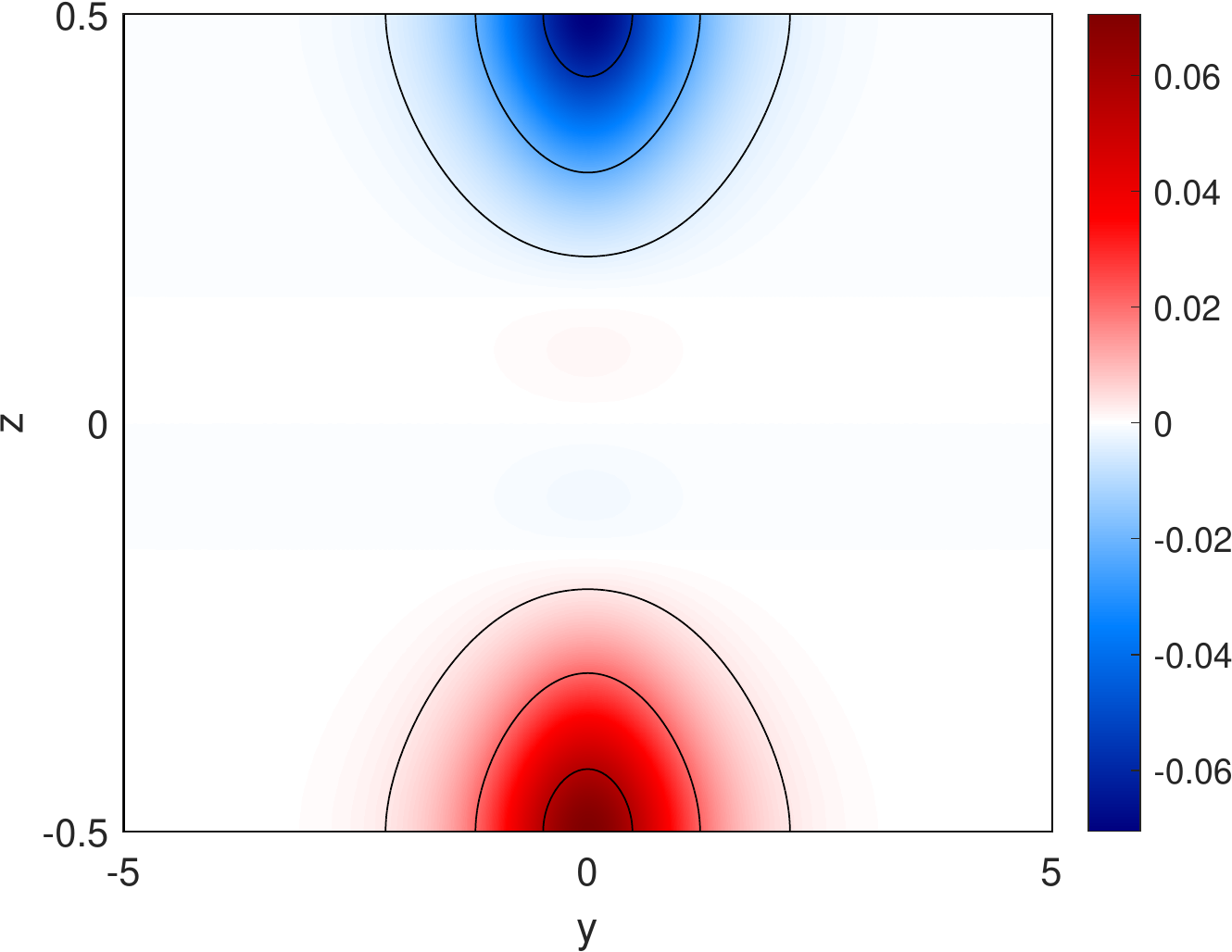}
	\caption{}
	\end{subfigure}
	\begin{subfigure}[b]{0.49\textwidth}
	\centering
	\includegraphics[trim={0 0 0 0},clip,width=\textwidth]{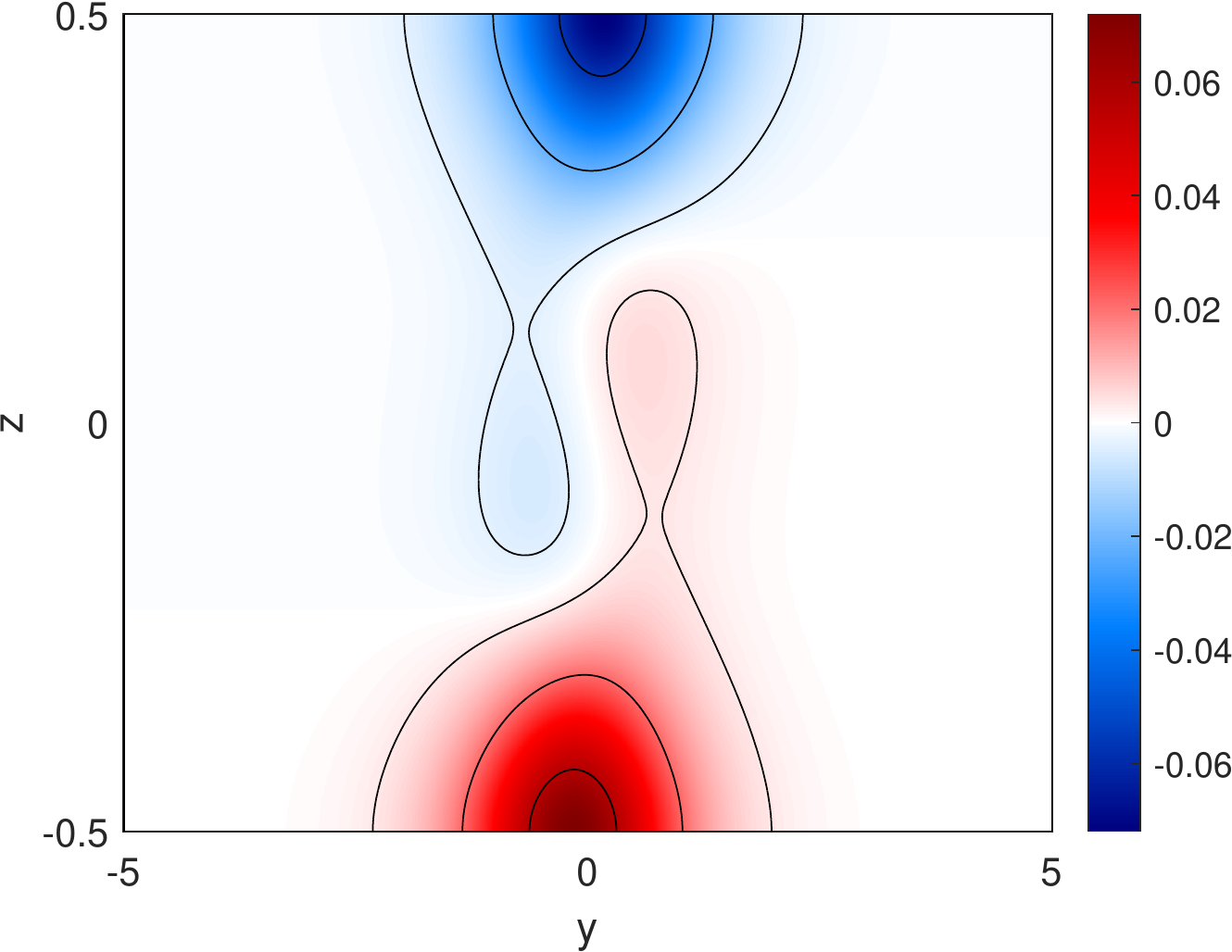}
	\caption{}
	\end{subfigure}
	\newline
	\begin{subfigure}[b]{0.49\textwidth}
	\centering
	\includegraphics[trim={0 0 0 0},clip,width=\textwidth]{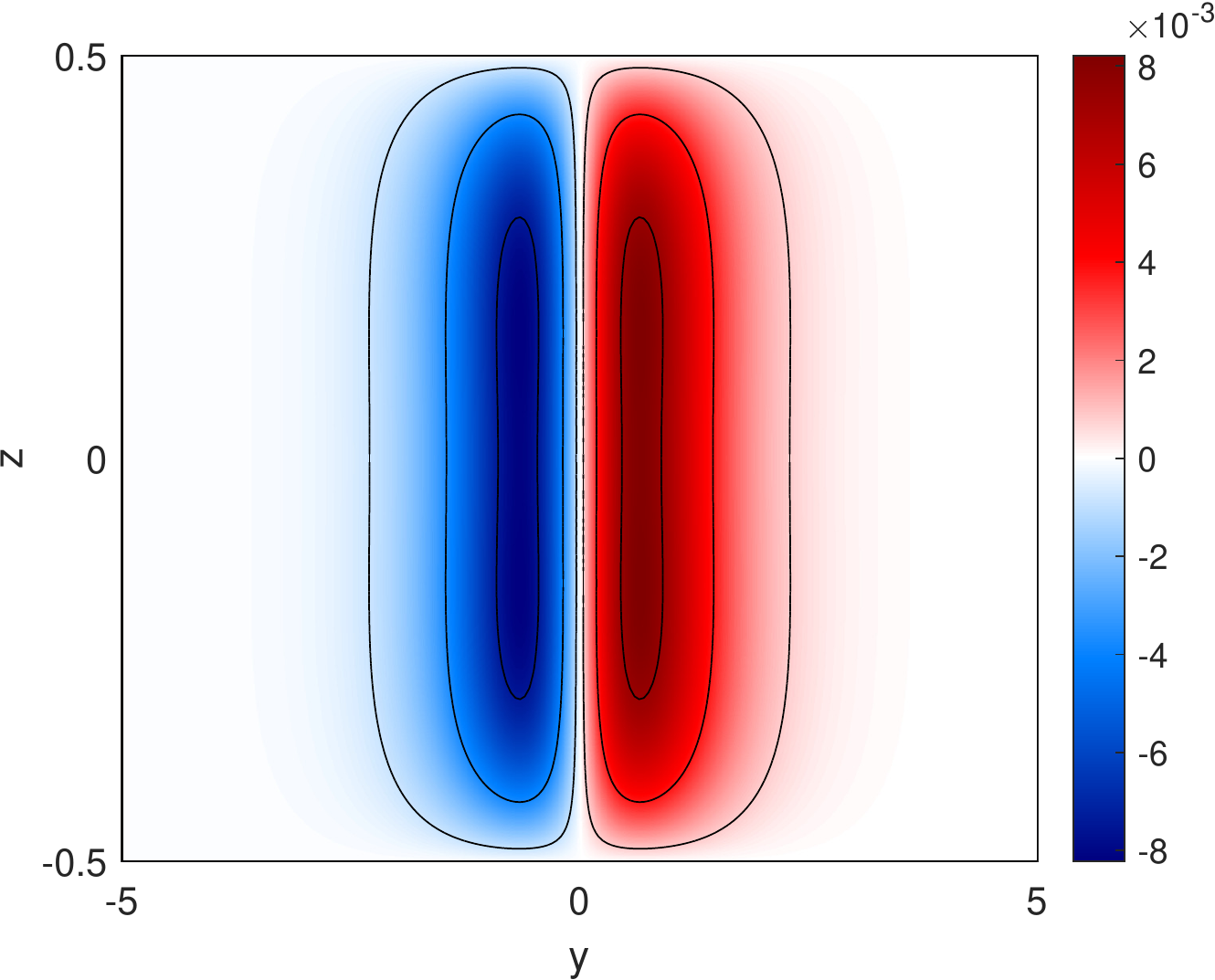}
	\caption{}
	\end{subfigure}
	\begin{subfigure}[b]{0.49\textwidth}
	\centering
	\includegraphics[trim={0 0 0 0},clip,width=\textwidth]{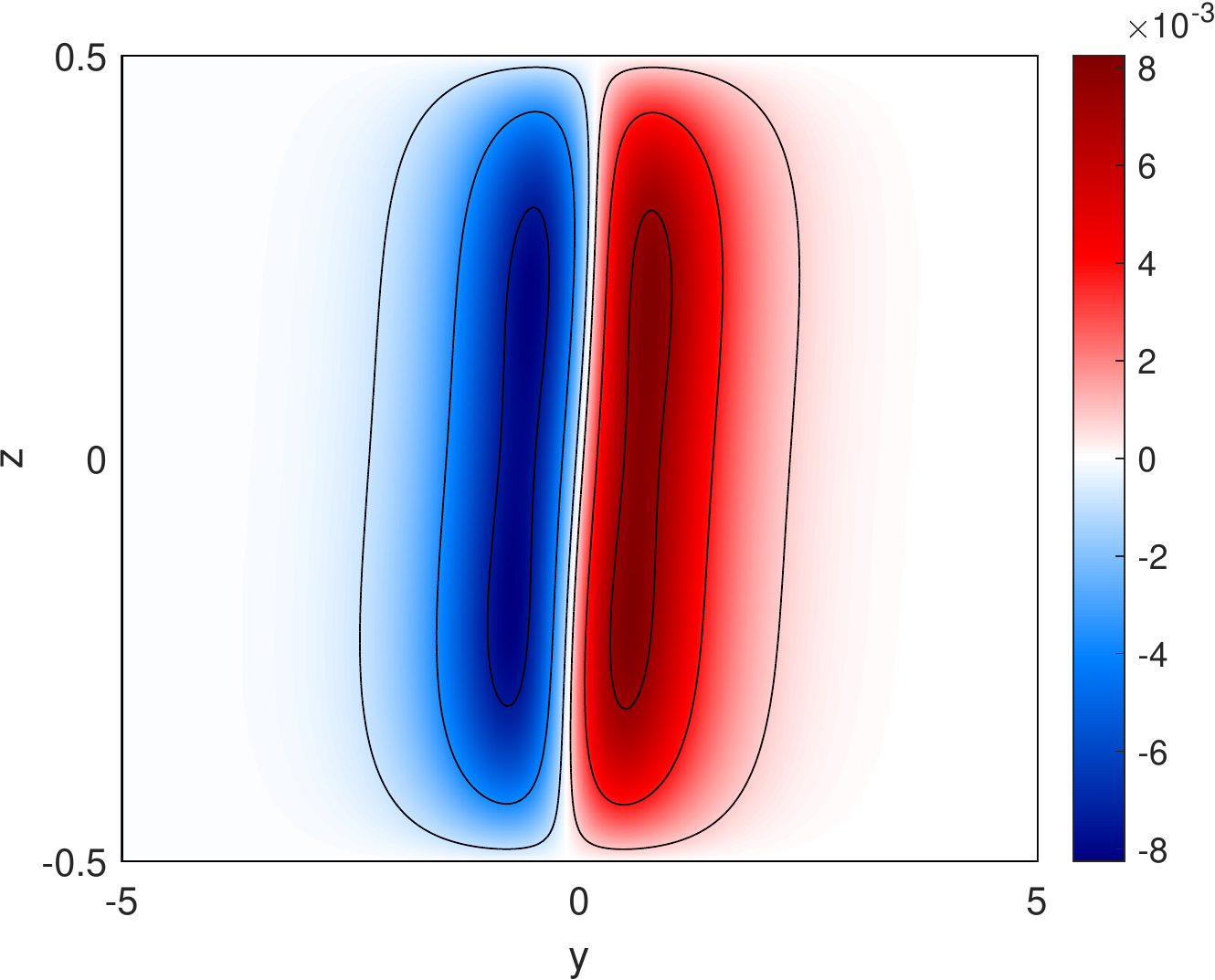}
	\caption{}
	\end{subfigure}
	\caption{Comparison between the TTW and modified TTW solutions for $\E = 0.01$ and $T = 1$. Solutions are shown correct to $O(\delta)$. (a) $\phi$ for $\delta = 0$, (b) $\phi$ for $\delta = 0.4$, (c) $v$ for $\delta = 0$, (d) $v$ for $\delta = 0.4$, (e) $w$ for $\delta = 0$, and (f) $w$ for $\delta = 0.4$.}
    \label{fig:phi_v_w}
\end{figure}

\cref{fig:phi_v_w} shows a comparison between the TTW solutions of \citet{CROWETAYLOR} (corresponding to $\delta = 0$) and the modified TTW solutions presented here with $\delta = 0.4$. Solutions are given correct to $O(\delta)$ using $\varphi = \varphi_0+\delta\,\varphi_1$ for a given field $\varphi$ and shown for $\E = 0.01$ and $T = 1$. The TTW solution consists of a flow from the high buoyancy side of the front to the low buoyancy side near the top surface and the opposite on the bottom surface. This results in upwelling on the high buoyancy side and downwelling on the low buoyancy side resulting in a anti-clockwise net circulation (shown by positive $\phi$). This behaviour was discussed in \citet{CROWETAYLOR} and is consistent with previous results and observations \citep{ELIASSEN,ORLANSKIROSS,MCWILLIAMS}. Non-traditional effects act to tilt the circulation cell and drive a flow in the centre of the layer. This flow may lead to a topological change in the structure of the circulation with a streamline in \cref{fig:phi_v_w}.(c) seen to split into two separate cells.

\begin{figure}
	\centering
	\begin{subfigure}[b]{0.49\textwidth}
	\centering
	\includegraphics[trim={0 0 0 0},clip,width=\textwidth]{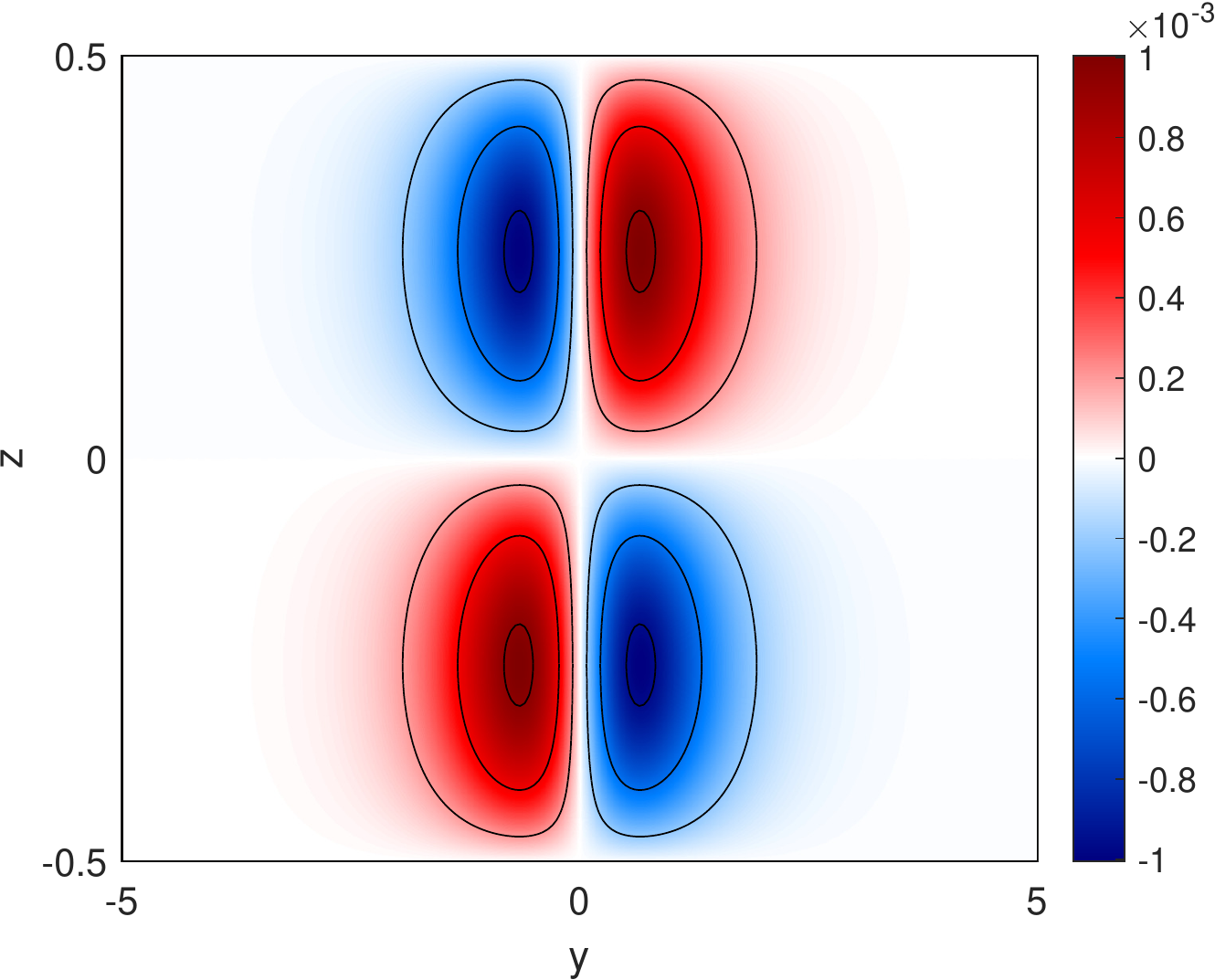}
	\caption{}
	\end{subfigure}
	\begin{subfigure}[b]{0.49\textwidth}
	\centering
	\includegraphics[trim={0 0 0 0},clip,width=\textwidth]{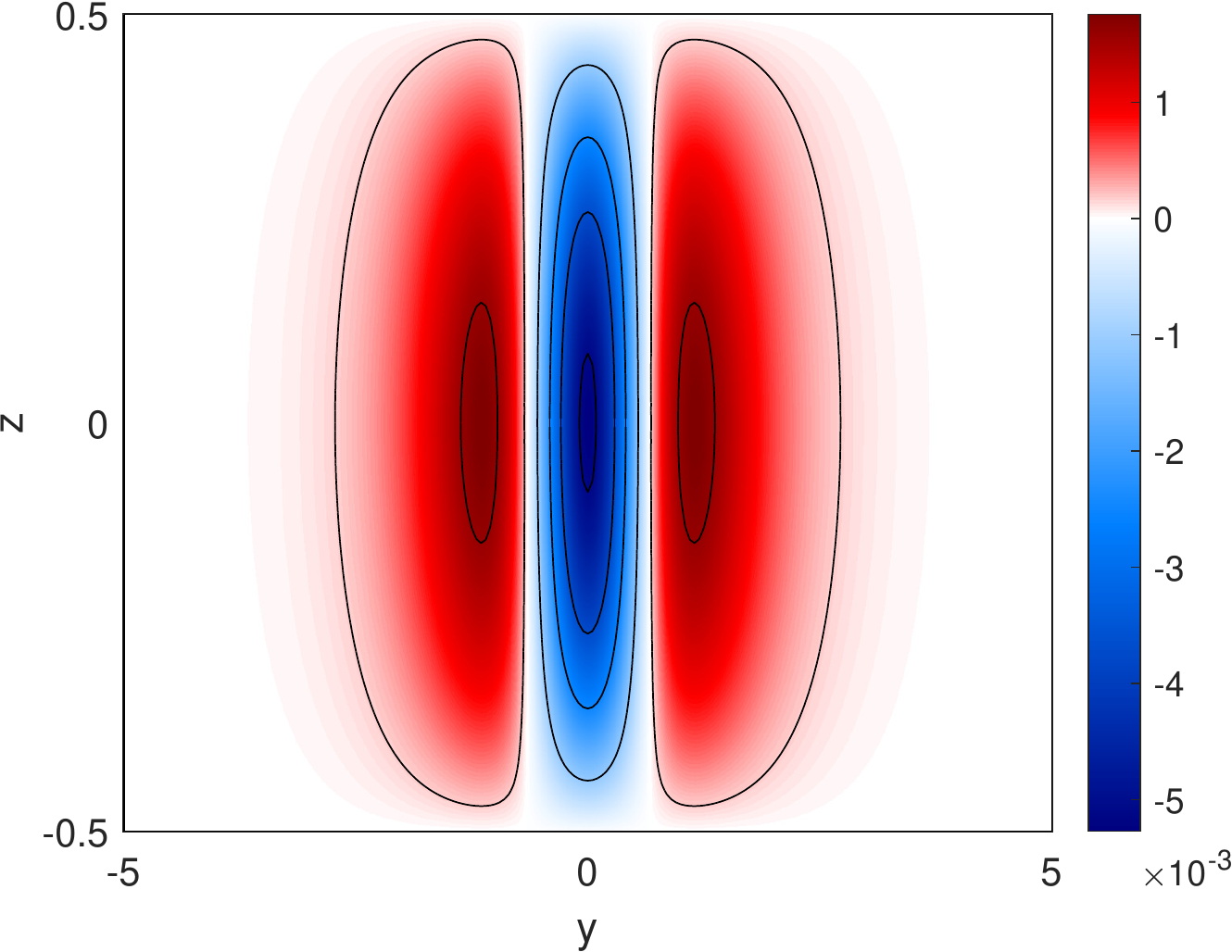}
	\caption{}
	\end{subfigure}
	\newline
	\begin{subfigure}[b]{0.49\textwidth}
	\centering
	\includegraphics[trim={0 0 0 0},clip,width=\textwidth]{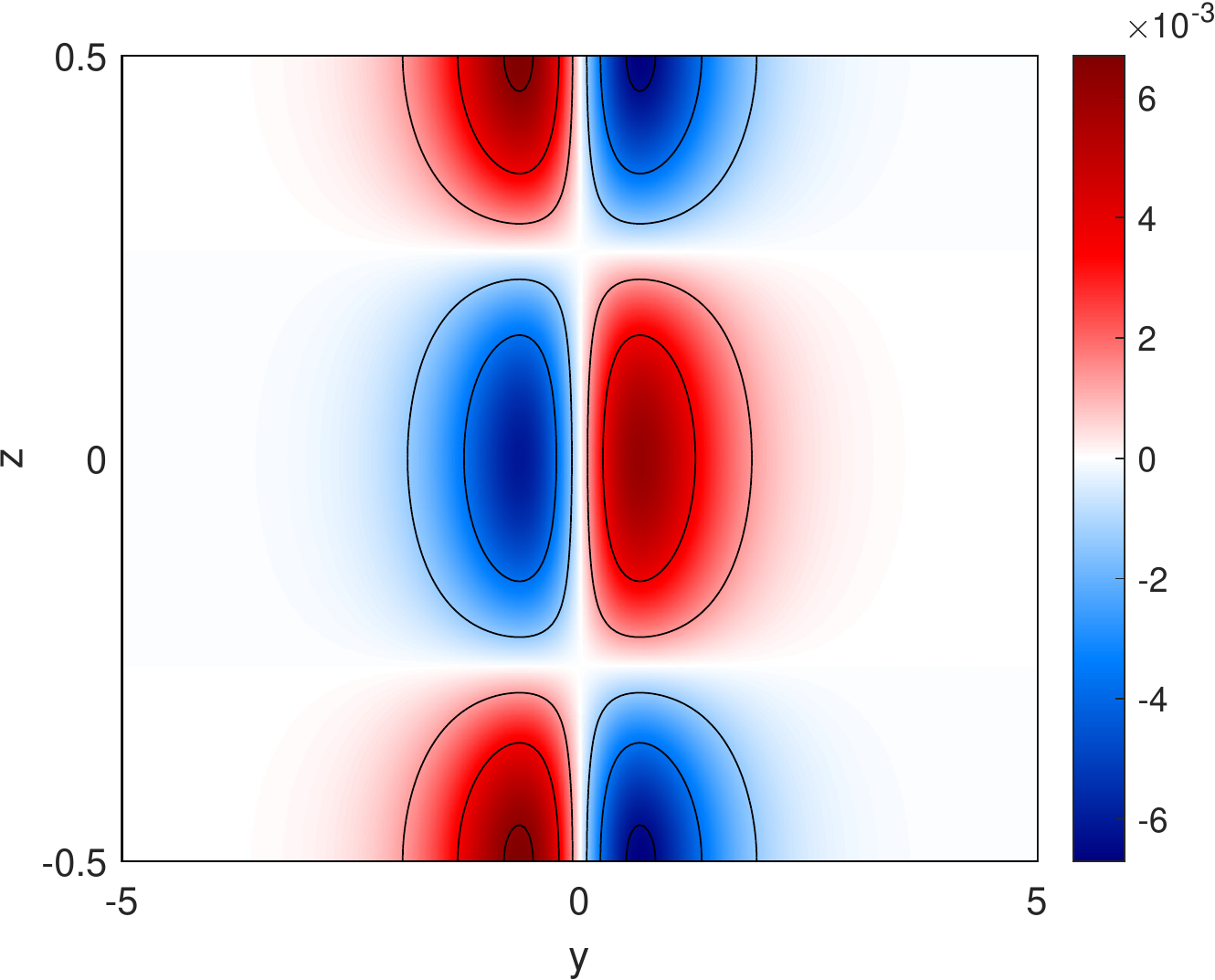}
	\caption{}
	\end{subfigure}
	\begin{subfigure}[b]{0.49\textwidth}
	\centering
	\includegraphics[trim={0 0 0 0},clip,width=\textwidth]{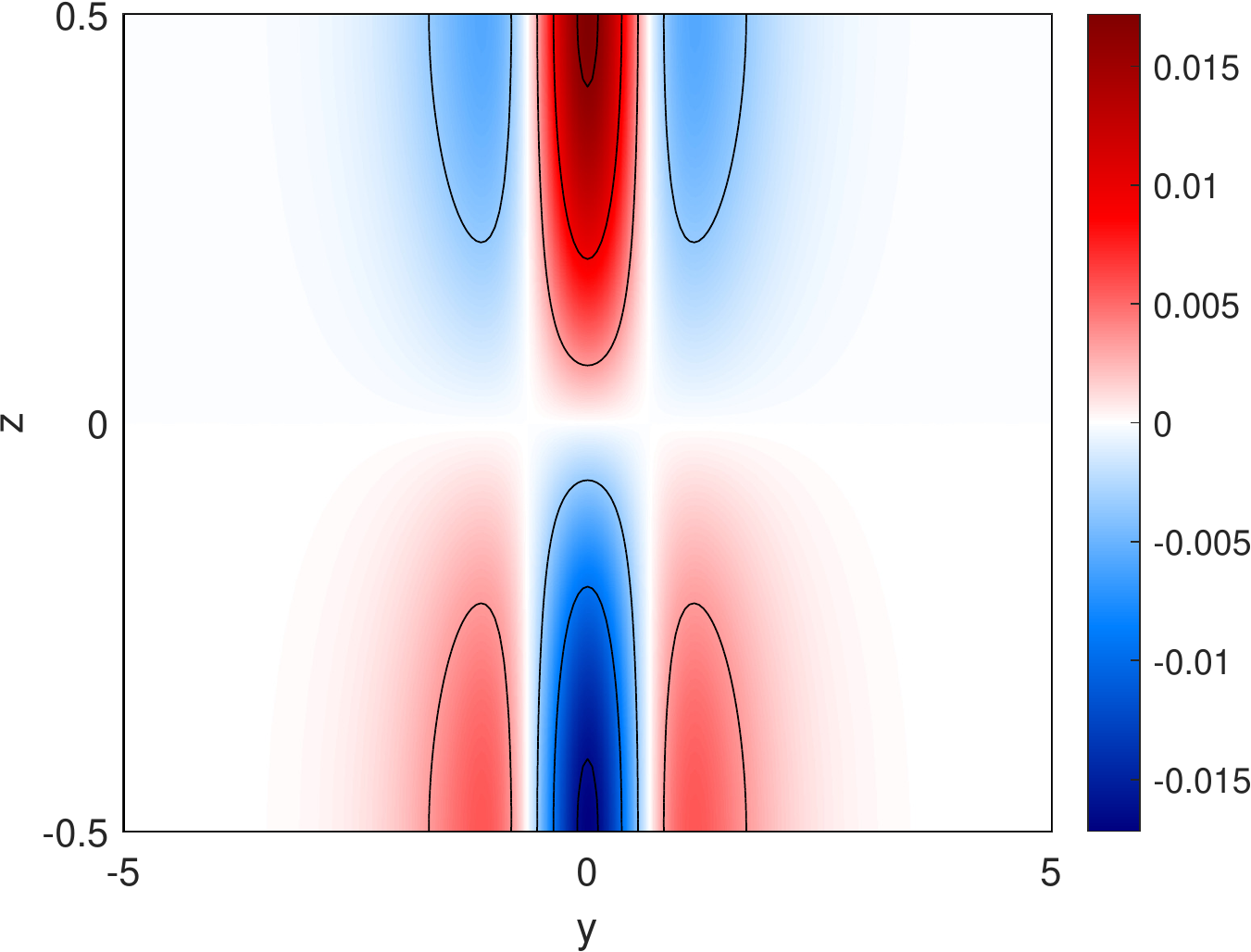}
	\caption{}
	\end{subfigure}
	\caption{(a) The first term of the $O(\delta)$ circulation streamfunction component, $\phi_1$, from \cref{eq:phi_sol}, (b) The second term of $\phi_1$ from \cref{eq:phi_sol}. Panels (c) and (d) show the cross-front velocity ($v$) components associated with the streamfunction components shown in panels (a) and (b) respectively. Results are shown for $\E = 0.1$ and $T = 1$.}
    \label{fig:phi_v}
\end{figure}

\cref{fig:phi_v} shows separately the two terms of $\phi_1$ from \cref{eq:phi_sol} for $\E = 0.1$ and $T = 1$. The associated cross-front velocities are also shown. The first term consists of four counter-rotating cells resulting in regions of convergence near the top and bottom boundaries and acting to tilt the leading order circulation cell. The second term consists of three counter rotating cells and results from the along-front jets modifying the vertical pressure gradient away from hydrostatic balance. As these jets grow, the second term of $\phi_1$ grows linearly with time for $T = O(1)$. Over very long timescales, it is predicted that these jets can become large enough to modify the absolute vertical vorticity in the frontal region. Therefore, while small in \cref{fig:phi_v_w}.(b), this circulation component may become large at late times leading to further topological changes in the structure of the frontal circulation.

Additionally, by depth-integrating the vertical velocities corresponding to the two terms in \cref{eq:phi_sol} the net vertical transport of fluid may be calculated. The first term depth-averages to zero so does not correspond to any vertical transport, instead this term describes a tilting of the circulation cell as noted above. The second term does, however, have a non-zero depth average which suggests that the circulation cells in \cref{fig:phi_v}.(b) may act to enhance the vertical exchange of tracers through the surface mixed layer.

The $O(\delta)$ cross-front velocities shown in \cref{fig:phi_v} contain regions of surface convergence. This velocity convergence can lead to a sharpening of surface buoyancy gradients resulting in frontogenesis \citep{HOSKINS,SHAKESPEARETAYLOR} and hence non-traditional effects may be frontogenetic. The asymptotic framework used here assumes $Ro \ll 1$ so this model is not strictly valid for studying frontogenesis where the Rossby number is typically order $1$. However, for $\Ro = O(1)$ the frontal sharpening predicted here will be an $O(\delta)$ effect, therefore, away from the equator, non-traditional effects are unlikely to be a dominant frontogenetic mechanism when compared to other mechanisms such as external strain, spontaneous adjustment and the secondary circulation induced by finite Rossby number effects \citep{HOSKINSBRETHERTON,BLUMEN,GULAETAL,MCWILLIAMS}. Non-traditional frontogenesis may be relevant in a small region around the equator where $\delta \geq O(1)$, though, since TTW is unlikely to be the dominant balance in this region, it is not possible to draw any conclusions from this analysis.

\subsection{The along-front flow}

The depth-dependent along-front velocity components are given by
\begin{equation}
\label{eq:u_example}
u_0' = \sqrt{\E}\,K(\z) \,\pder{b_0}{y}, \quad u_1' = - \E\,B(\z)\,\pder{^2b_0}{y^2} -\sqrt{\E}\, K(\z)\, \pder{^2\psi_0}{y^2},
\end{equation}
were the two terms of $u_1'$ arise from the modified horizontal momentum balance and the modified hydrostatic balance similarly to the terms of $\phi_1$ in \cref{eq:phi_sol}.

\begin{figure}
	\centering
	\begin{subfigure}[b]{0.49\textwidth}
	\centering
	\includegraphics[trim={0 0 0 0},clip,width=\textwidth]{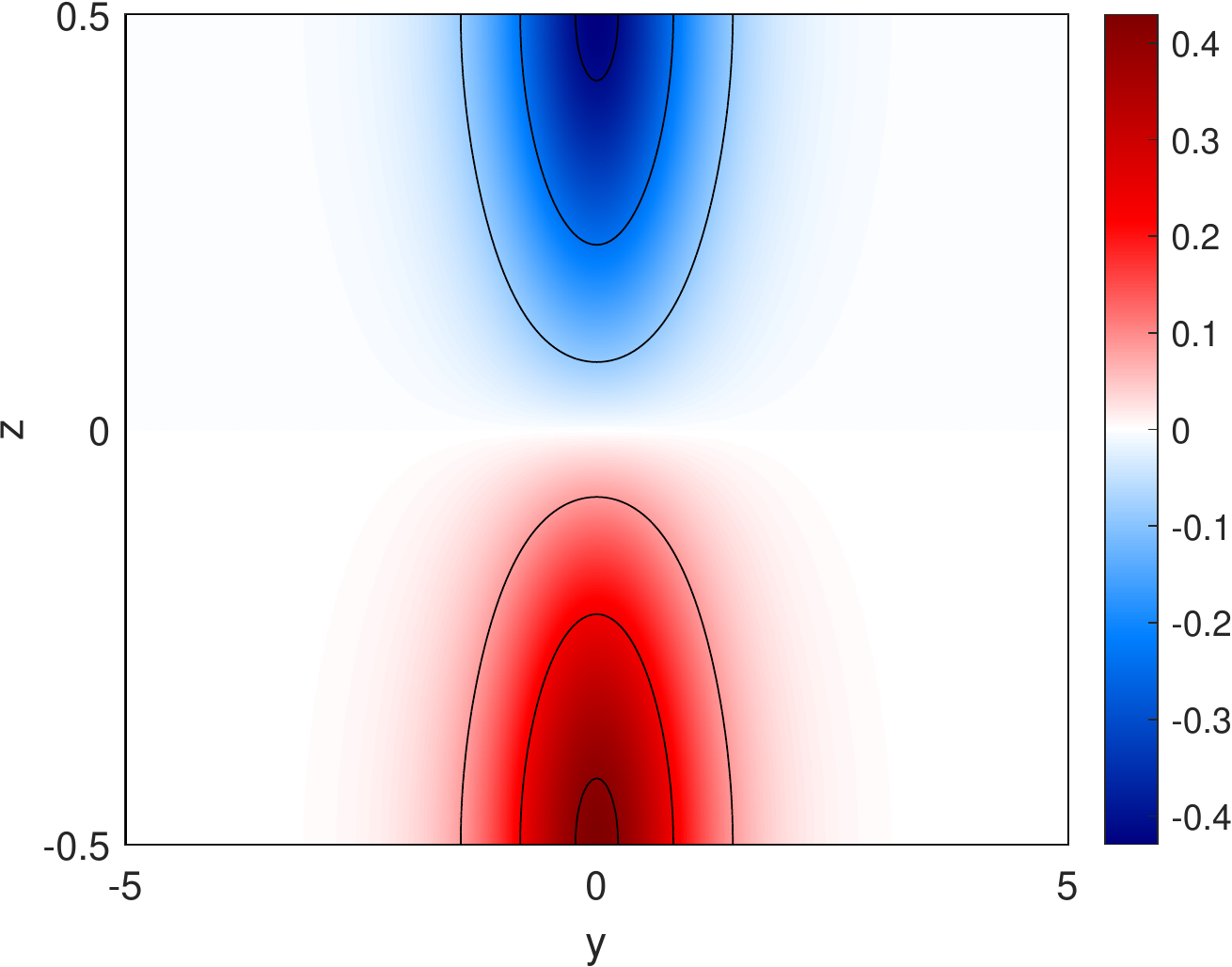}
	\caption{}
	\end{subfigure}
	\begin{subfigure}[b]{0.49\textwidth}
	\centering
	\includegraphics[trim={0 0 0 0},clip,width=\textwidth]{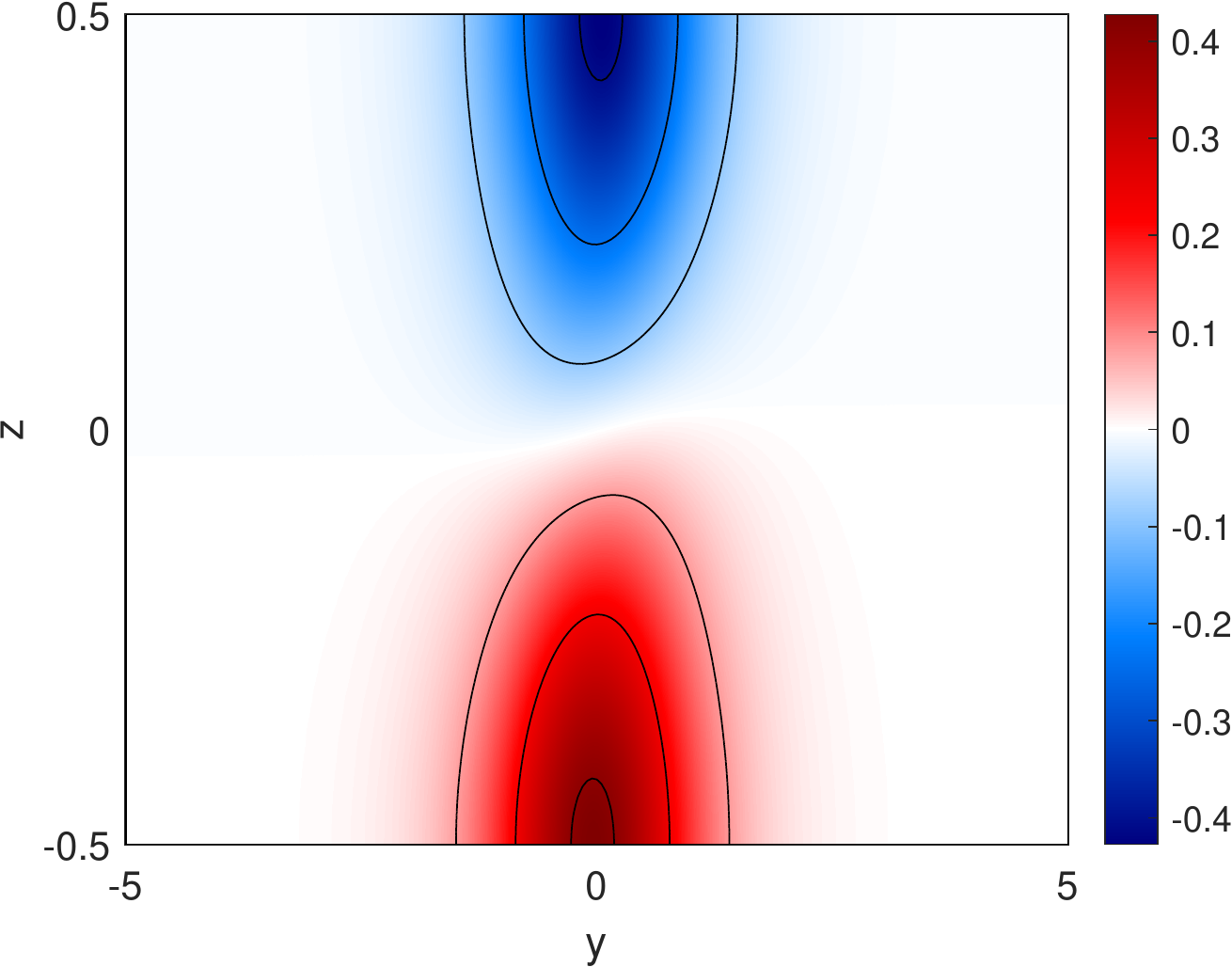}
	\caption{}
	\end{subfigure}
	\caption{Comparison of the along-front velocities for the TTW and modified TTW systems. Parameters are $\E = 0.01$ and $T = 1$ and solutions are given correct to $O(\delta)$. (a) $u'$ for $\delta = 0$, (b) $u'$ for $\delta = 0.4$,}
    \label{fig:u}
\end{figure}

\cref{fig:u} shows the depth-dependent along-front velocity correct to $O(\delta)$ for the cases of $\delta = 0$ and $\delta = 0.4$ with $T = 1$. The along-front flow is dominated by a thermal wind shear modified by vertical mixing \citep{CROWETAYLOR} and non-traditional effects are seen to be small. Therefore, the most significant effect of non-traditional rotation on the along-front flow is the development of the depth-independent jets shown in \cref{fig:psi} though if the jets become large, significant modification of the depth-dependent flow may occur through the second term of \cref{eq:u_example}.

\subsection{Buoyancy and stratification}

\begin{figure}
	\centering
	\begin{subfigure}[b]{0.49\textwidth}
	\centering
	\includegraphics[trim={0 0 0 0},clip,width=\textwidth]{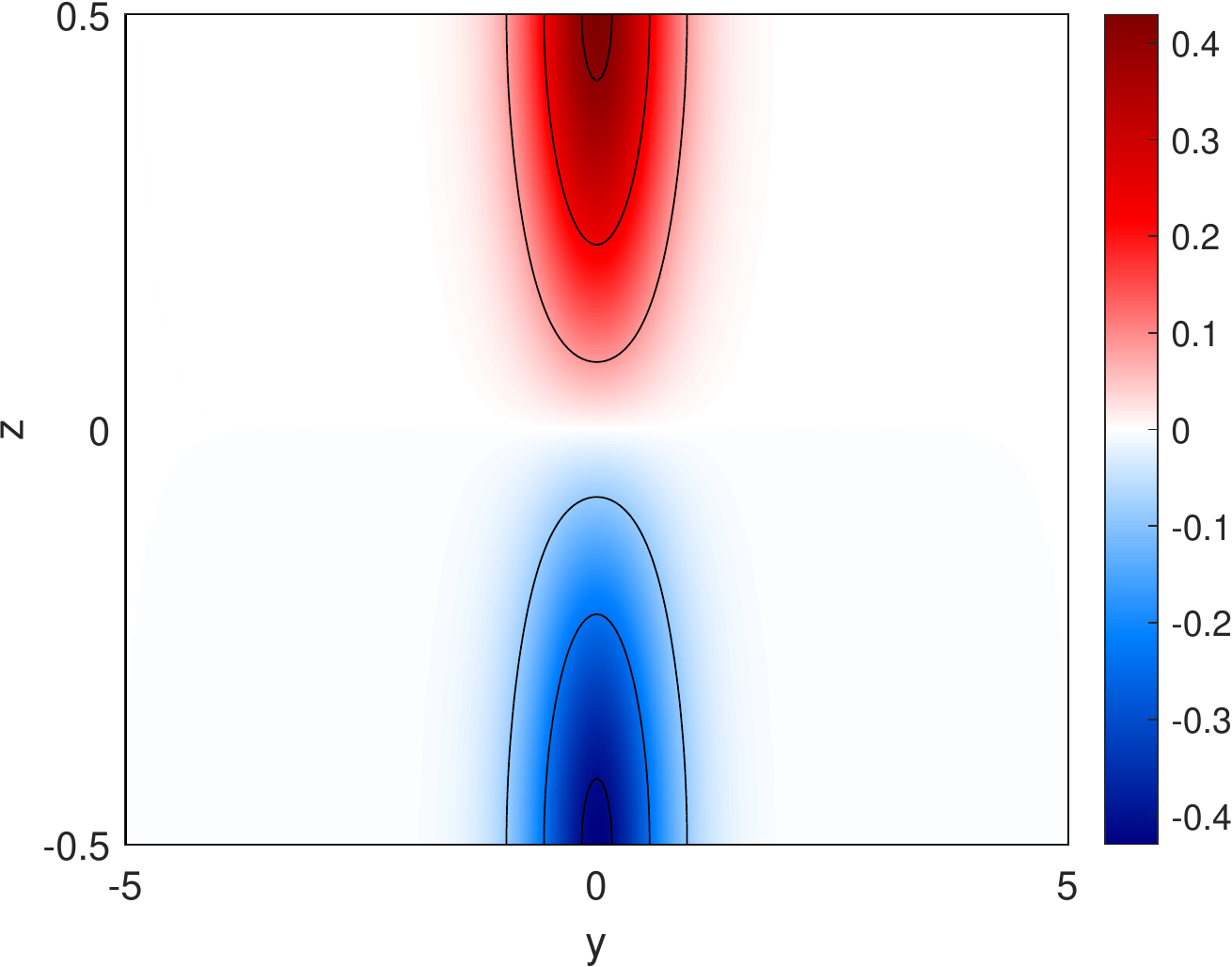}
	\caption{}
	\end{subfigure}
	\begin{subfigure}[b]{0.49\textwidth}
	\centering
	\includegraphics[trim={0 0 0 0},clip,width=\textwidth]{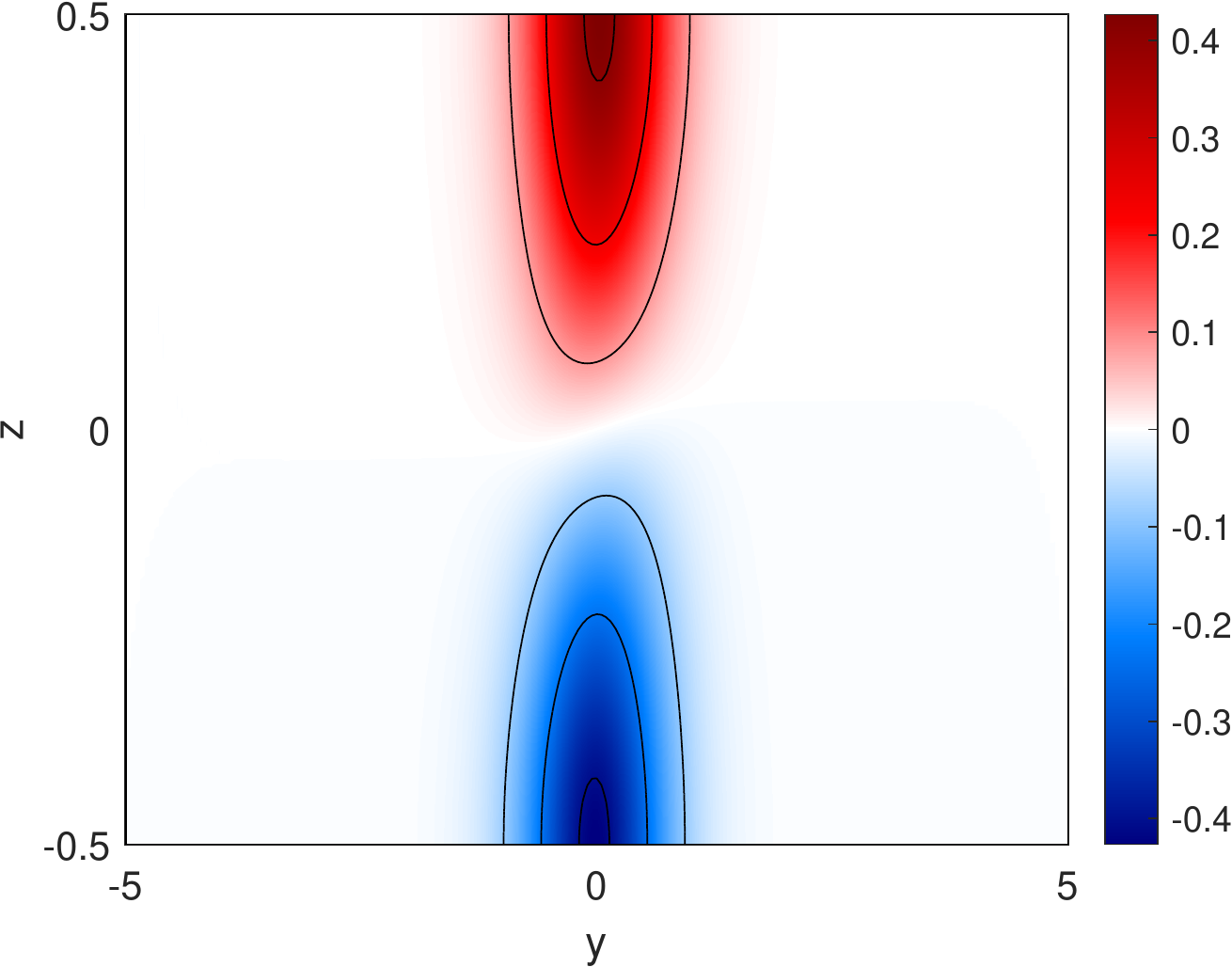}
	\caption{}
	\end{subfigure}
	\newline
	\begin{subfigure}[b]{0.49\textwidth}
	\centering
	\includegraphics[trim={0 0 0 0},clip,width=\textwidth]{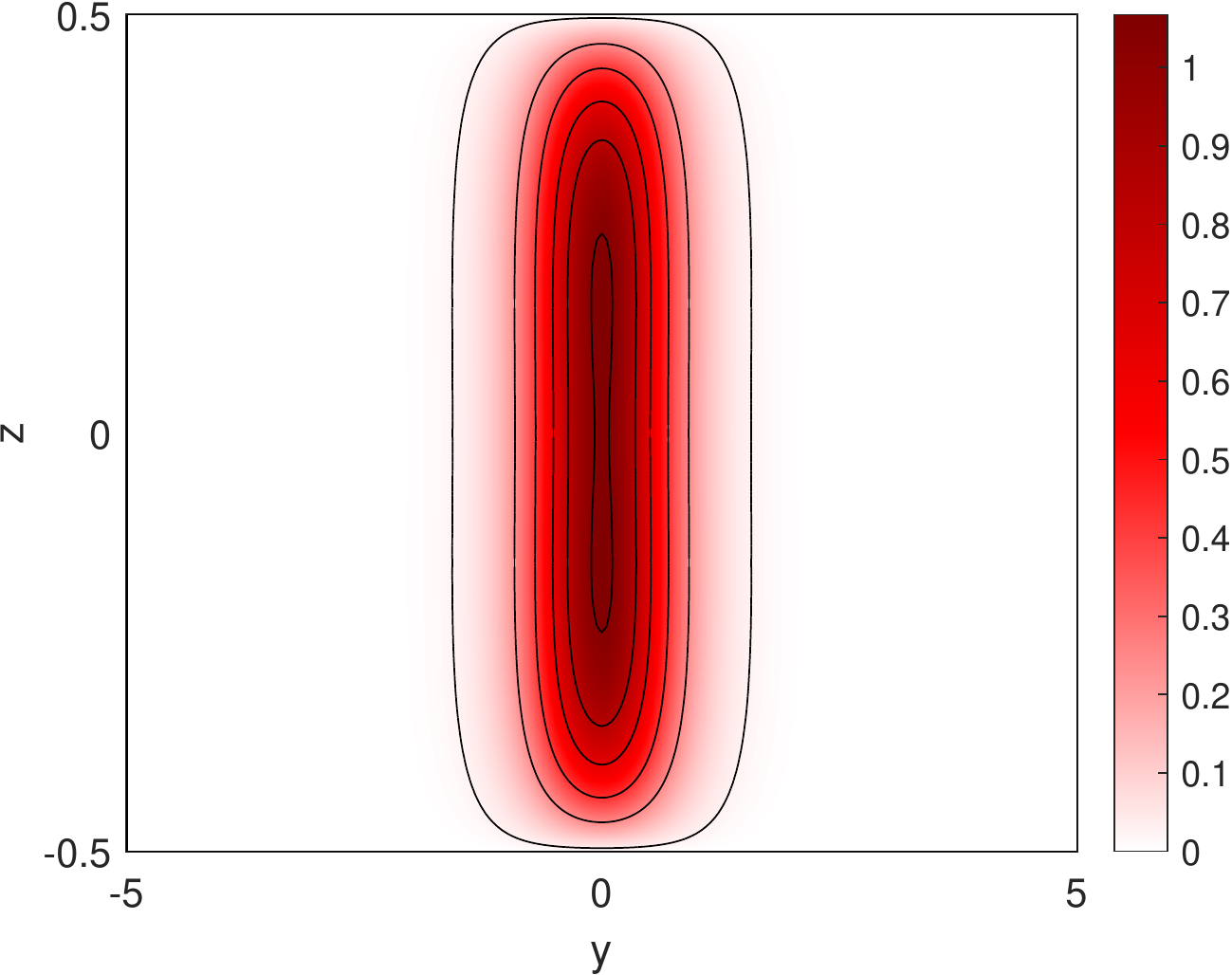}
	\caption{}
	\end{subfigure}
	\begin{subfigure}[b]{0.49\textwidth}
	\centering
	\includegraphics[trim={0 0 0 0},clip,width=\textwidth]{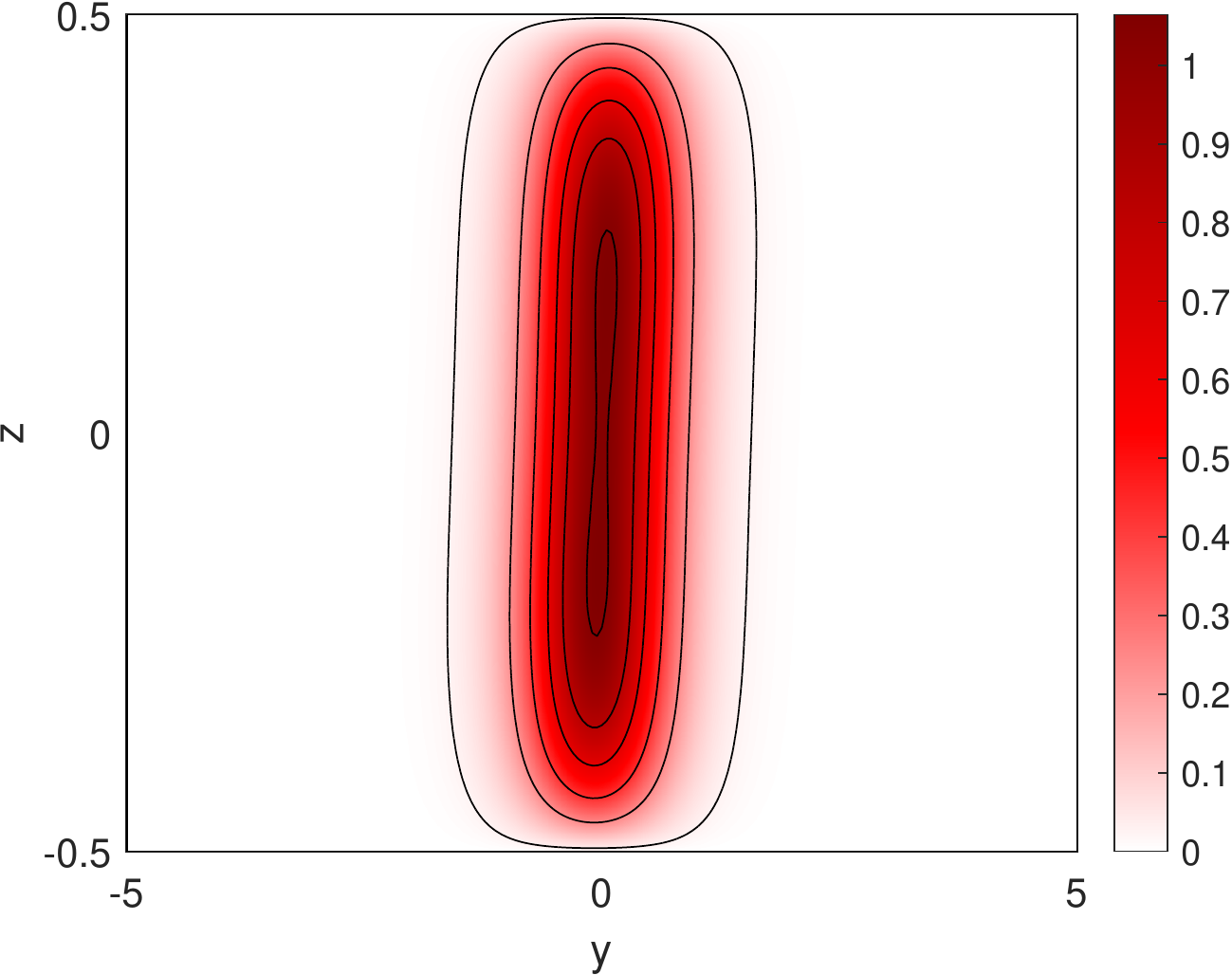}
	\caption{}
	\end{subfigure}
	\caption{Comparison between the TTW and modified TTW solutions for $\E = 0.01$ and $T = 1$. Solutions are shown correct to $O(\delta)$. (a) $b'/(\Ro\,\Pr)$ for $\delta = 0$, (b) $b'/(\Ro\,\Pr)$ for $\delta = 0.4$, (c) $N^2/(\Ro\,\Pr)$ for $\delta = 0$, and (d) $N^2/(\Ro\,\Pr)$ for $\delta = 0.4$.}
    \label{fig:b_N2}
\end{figure}

As noted above, the depth averaged buoyancy remains equal to $b_0$ for the simple case of $b_0 = \tanh y$. However, the vertical structure of the buoyancy field and the associated stratification are determined by the frontal circulation so are still affected by non-traditional rotation. These terms appear at orders $O(\delta^2)$ and $O(\delta^3)$ and from \cref{eq:bp_summ} are given by
\begin{equation}
\label{eq:bp_example}
b_2 = -\mathcal{R}\Pr\sqrt{\E}\,K(\z)\left(\pder{b_0}{y}\right)^2\!\!,\quad\! b_3 = \mathcal{R}\Pr\left(\! \E\,D_1(\z)\pder{b_0}{y}\pder{^2b_0}{y^2} \!+\! \sqrt{\E}\,K(\z) \pder{b_0}{y}\pder{^2\psi_0}{y^2}\!\right)\!.\!\!
\end{equation}
The lowest order buoyancy term with vertical structure, $b_2$, describes the stratification maintained by an advection-diffusion balance between the advection of buoyancy by the leading order circulation, $\phi_0$, and the vertical mixing of buoyancy \citep{CROWETAYLOR}. A similar balance occurs at $O(\delta^3)$ so the first (second) term of $b_3$ in \cref{eq:bp_example} describes the stratification maintained by the first (second) term of $\phi_1$.

\cref{fig:b_N2} shows the depth-dependent buoyancy, $b'$, and vertical stratification, $N^2 = \pderline{b}{z}$, correct to $O(\delta^3)$ for $\E = 0.01$ and $T = 1$. Solutions are shown for $\delta = 0$ and $\delta = 0.4$. Since $b'$ and $N^2$ are linear in $\Ro\,\Pr$ through the factor of $\mathcal{R}\,\delta^2\,Pr$, results are plotted for $b'/(\Ro\,\Pr)$ and $N^2/(\Ro\,\Pr)$ to remove this dependence. The advection-diffusion balance is seen to drive a stable restratification of the front and modification by non-traditional effects is small unless $\psi_0$ becomes large.

\begin{figure}
	\centering
	\begin{subfigure}[b]{0.49\textwidth}
	\centering
	\includegraphics[trim={0 0 0 0},clip,width=\textwidth]{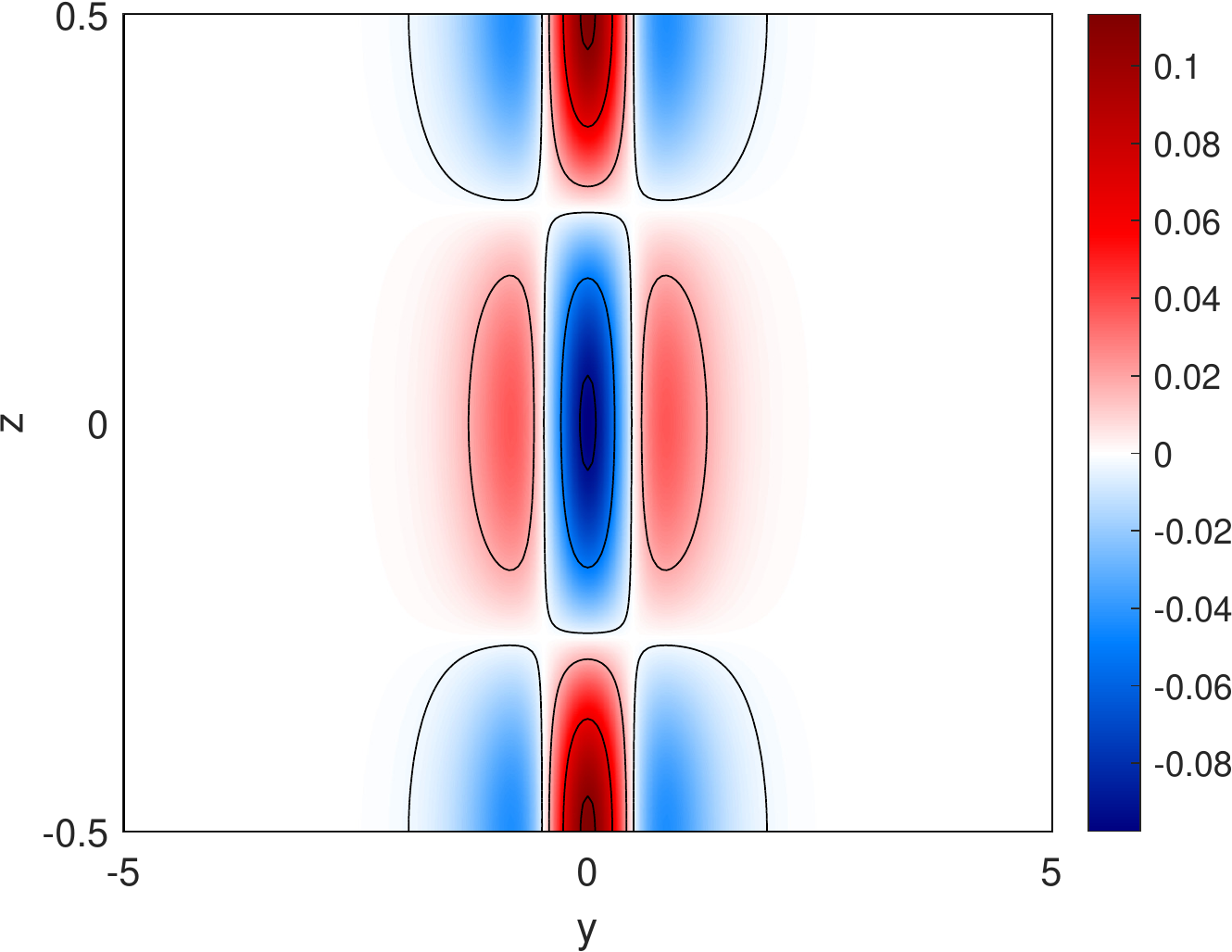}
	\caption{}
	\end{subfigure}
	\begin{subfigure}[b]{0.49\textwidth}
	\centering
	\includegraphics[trim={0 0 0 0},clip,width=\textwidth]{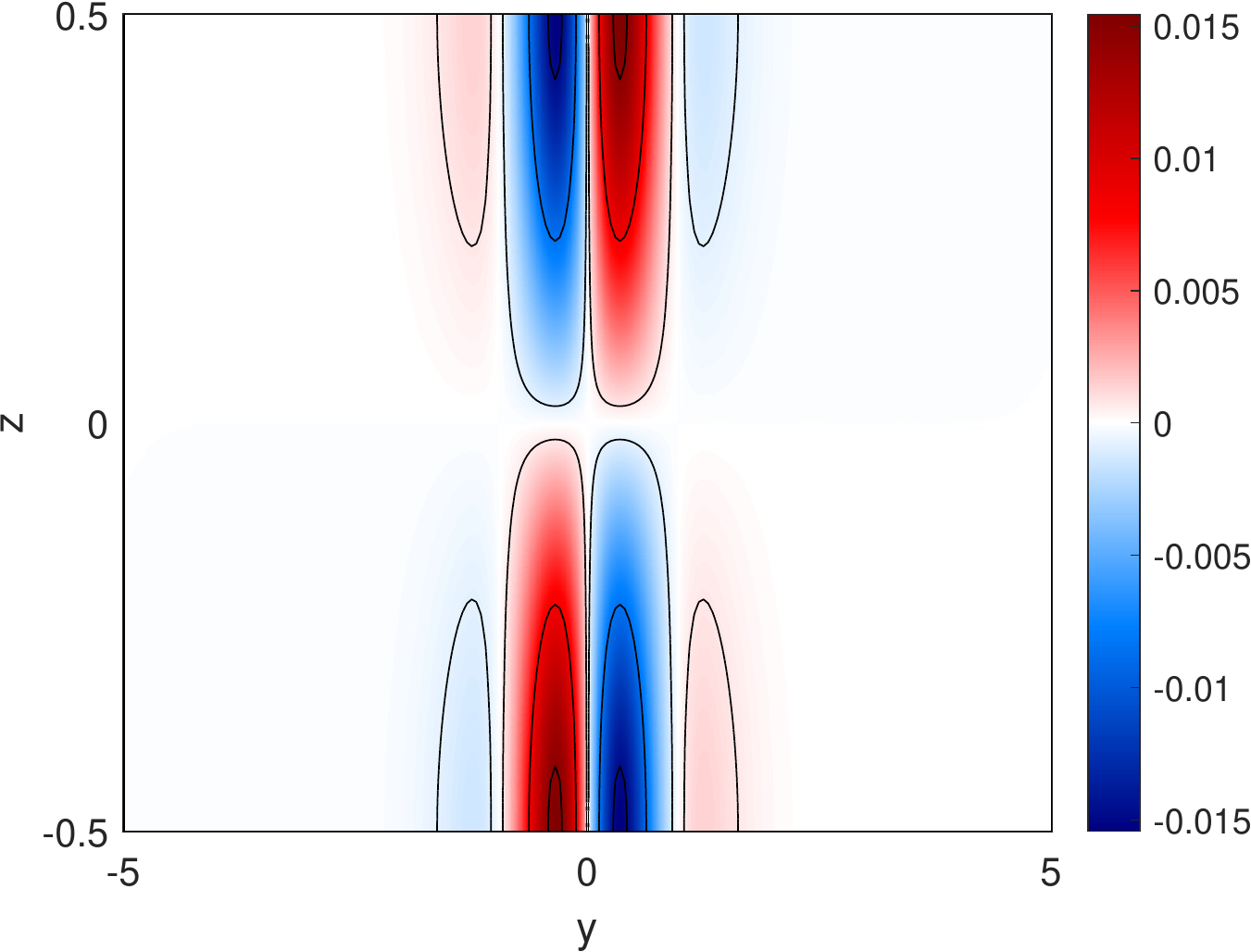}
	\caption{}
	\end{subfigure}
	\caption{(a) The first term of $M_3^2$ from \cref{eq:M2_example}. (b) The second term of $M_3^2$ from \cref{eq:M2_example}. Results are shown for $\E = 0.01$, $\mathcal{R}\,\Pr = 1$, and $T = 1$.}
	\label{fig:M2}
\end{figure}

From \cref{eq:bp_example} the order $O(\delta^3)$ horizontal buoyancy gradient may be calculated as
\begin{equation}
\label{eq:M2_example}
{M^2_3} = \pder{b_3}{y} = \mathcal{R}\Pr\left( \E\,D_1(\z)\pder{}{y}\!\left[\pder{b_0}{y}\pder{^2b_0}{y^2}\right] + \sqrt{\E}\,K(\z) \pder{}{y}\!\left[\pder{b_0}{y}\pder{^2\psi_0}{y^2}\right]\right).
\end{equation}
The two terms of \cref{eq:M2_example} are plotted in \cref{fig:M2} for $\E = 0.01$, $\mathcal{R}\,\Pr = 1$ and $T = 1$. Regions of positive horizontal buoyancy gradient are observed for both terms in $M_3^2$,  these regions correspond to frontal sharpening due to the cross-front velocity convergence seen in \cref{fig:phi_v}.

\subsection{Shear dispersive spreading}

For $b_0 = b_0(y)$, \cref{eq:b_0_spread} becomes
\begin{equation}
\label{eq:b_0_example}
\pder{b_0}{t} = \Ro^2\,\Pr\, \pder{}{y}\left[c_3 \left(\pder{b_0}{y}\right)^3+\delta\,c_4 \left(\pder{b_0}{y}\right)^2\!\pder{^2b_0}{y^2}\right],
\end{equation}
for
\begin{equation}
c_3(\E) = \E\,\da{K'^2} \quad \textrm{and} \quad c_4(\E) = 2 \sqrt{\E^3}\,\da{AK}.
\end{equation}
This equation is derived using the $\psi_0$ independent terms from $b'_3$ and $\textbf{u}'_{H1}$ and corresponds to the case of weak vorticity generation, $\psi_0 = 0$. Over the long timescale of shear dispersive spreading, $t=O(1/\delta^4)$, significant vorticity generation is expected. However, the case of $\psi_0 = 0$ is considered here to isolate the effect of the $\psi$ independent terms.

As $\delta \to 0$, \cref{eq:b_0_example} reduces to the result of \citet{CROWETAYLOR} where the front approaches a self-similar solution and spreads as $y \sim t^{1/4}$. This self-similar solution is odd in $y$ hence the high buoyancy and low buoyancy sides evolve in the same way. However, the $c_4$ term breaks this $y$ symmetry due to an odd number of $y$ derivatives so both sides are expected to evolve differently for non-zero $\delta$. To test this prediction \cref{eq:b_0_example} is solved numerically using the Dedalus framework \citep{BurnsVOLB20}. The units of time are re-scaled such that $\Ro^2\Pr\,c_3 = 1$ leaving $r = \delta\,c_4/c_3$ as the only free parameter. Simulations are run for $r = 0$ and $r = 0.2$ and initialised using the profile $b_0(t = 0) = \tanh y$. Sixth-order hyperdiffusion with a hyperdiffusivity of $\nu_6 = 3\times 10^{-9}$ is included for numerical stability and simulations are run until $\Ro^2\Pr\,c_3 \,t = 10$ using a third order implicit-explicit Runge-Kutta scheme and a domain of $y \in [-5,5]$ with $N_y = 256$ grid-points.

\begin{figure}
	\centering
	\begin{subfigure}[b]{0.49\textwidth}
	\centering
	\includegraphics[trim={0 0 0 0},clip,width=\textwidth]{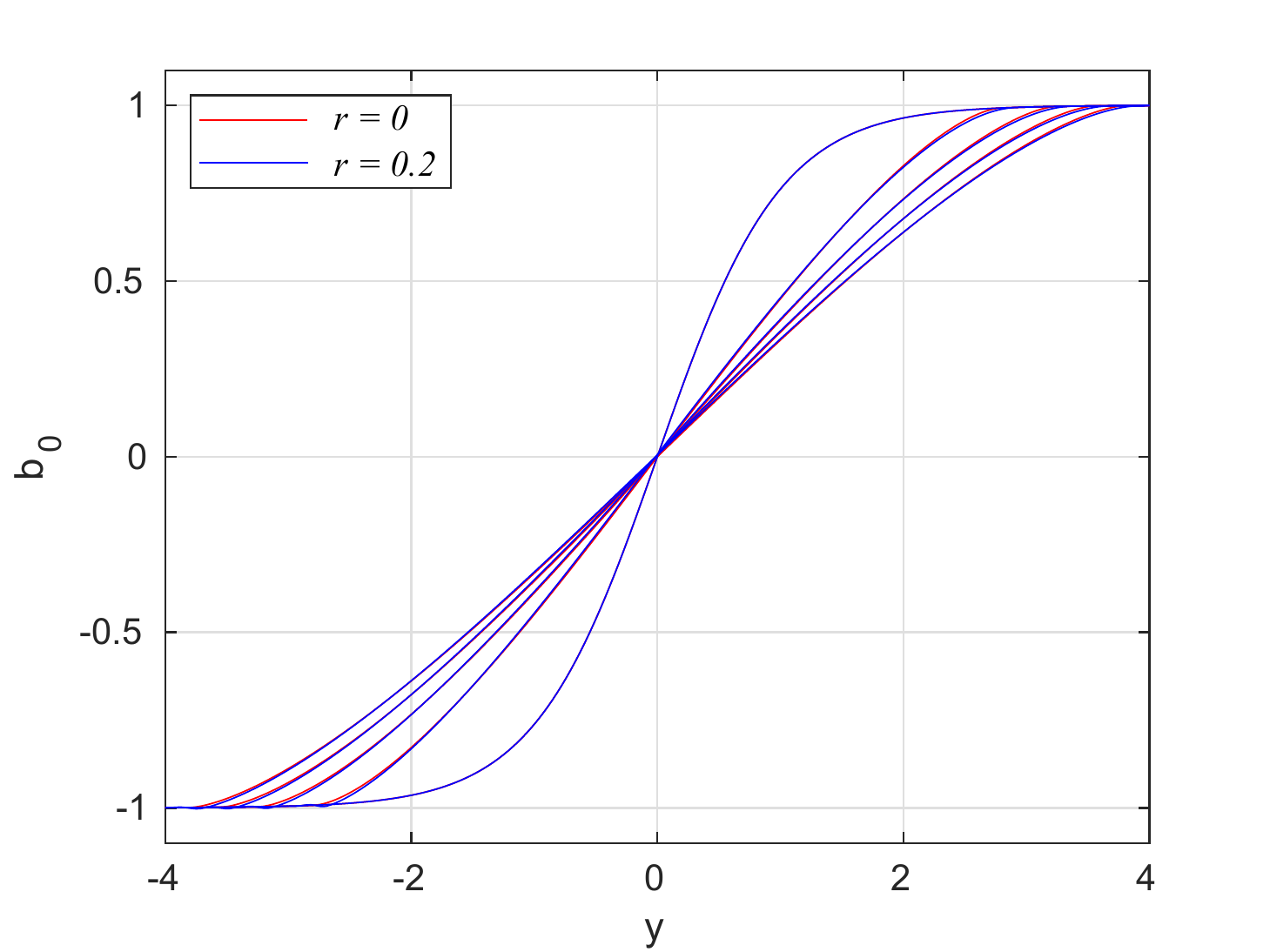}
	\caption{}
	\end{subfigure}
	\begin{subfigure}[b]{0.49\textwidth}
	\centering
	\includegraphics[trim={0 0 0 0},clip,width=\textwidth]{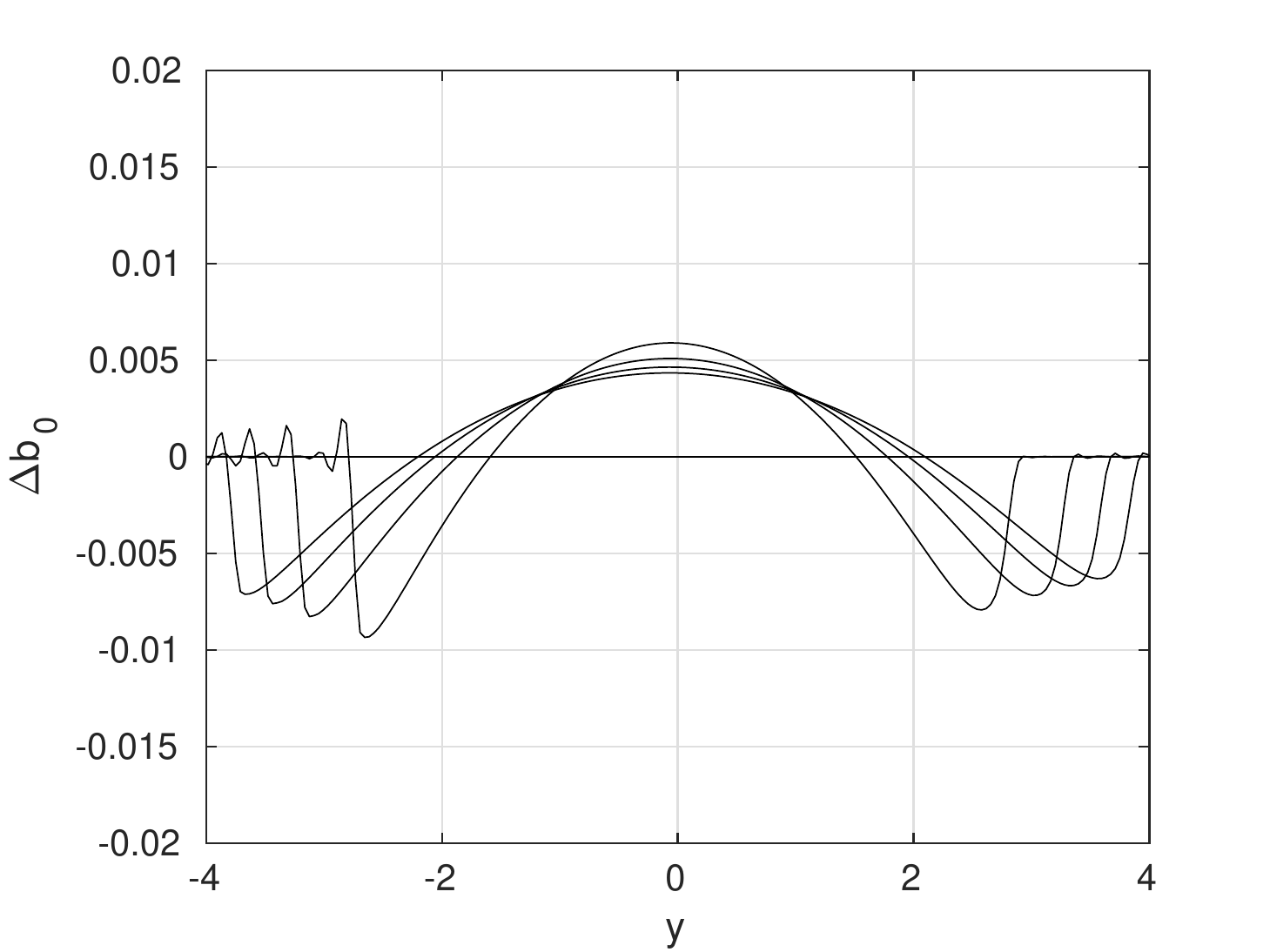}
	\caption{}
	\end{subfigure}
	\caption{(a) The buoyancy field, $b_0$, from numerically solving \cref{eq:b_0_example}. Solutions are calculated for $r = \delta\,c_4/c_3 = 0$ (red) and $r = 0.2$ (blue) and shown for $\Ro^2\Pr\,c_3\,t \in [0, 2.5, 5, 7.5, 10]$ with steeper solutions corresponding to earlier times. (b) The difference in $b_0$ between the $r = 0$ results and the $r = 0.2$ results for the same time values as (a).}
	\label{fig:b_spread}
\end{figure}

\cref{fig:b_spread} shows the numerical solutions for $b_0$ for $r = 0$ and $r = 0.2$ at a range of times. The difference between these two solutions is plotted in panel (b) allowing the expected asymmetry due to non-traditional effects to be observed. The effect of the non-traditional term is found to be small and of greatest importance near the frontal edges where the curvature, $\pderline{^2b_0}{y^2}$, is large. Therefore it is expected that non-traditional effects will not play a significant role in the shear dispersive spreading of a front without the inclusion of strong vorticity generation.

Over long times, a large amount of vorticity is expected to be generated by both vertical mixing \citep{CROWETAYLOR3} and non-traditional effects. This vorticity manifests as depth-independent jets and can act to modify the total vorticity of the system. If this vorticity is sufficiently strong, local vorticity terms can appear in the leading order turbulent thermal wind balance. These terms will modify the Coriolis force, resulting in a modified depth-dependent velocity and hence a modified circulation. Therefore it is predicted that the generation of vorticity will play a more important role in the evolution of $b_0$ than the $\psi_0$ independent circulation considered here. The effects of large $\psi_0$ could be considered using the approach of \citet{CROWETAYLOR3} however such solutions are expected to be complicated and provide no new insight into the problem so will not be considered here.

\section{The finite Rossby number regime}

Throughout we have taken $\Ro = O(\delta^2)$. This assumption is predominantly for mathematical convenience as it results in linear equations for the velocity at the first two orders in $\delta$. However, frontal systems in which non-traditional effects are important are unlikely to have small Rossby numbers. Here, typical frontal parameters are discussed and numerical simulations are presented for parameters outside of the regime considered above.

\subsection{Typical frontal parameters}

The small parameter describing non-traditional effects is the ratio of the vertical and horizontal components of the rotation vector scaled by the aspect ratio and is given by
\begin{equation}
\delta = \frac{H}{L} \frac{1}{\tan \theta},
\end{equation}
for latitude $\theta$, layer depth $H$ and typical frontal width $L$. The requirement that $\Ro = O(\delta^2)$ therefore implies that
\begin{equation}
B \sim H \tw{f}^2,
\end{equation}
where $B$ is the typical buoyancy difference across the front. Taking typical values of $\tw{f} \approx 10^-4 \,\textrm{s}^{-1}$ and $H = 100\,\textrm{m}$ gives a buoyancy difference which is much smaller than the typical values of $B\approx 10^{-4}\,\textrm{ms}^{-2}$. Therefore, for this asymptotic regime to hold, the frontal velocities and hence the Rossby number would have to be much smaller than would be expected physically.

To determine more physical values of the relevant parameters, note that non-traditional effects are most important in the tropical and subtropical regions where $\tan \theta < 1$. Here, fronts with small horizontal scales, $L \sim 1 \, \textrm{km}$, may have order $1$ values of $\delta$; however, the Rossby number would also be order $1$ for these small scale fronts. This case of $\Ro = O(1)$ and $\delta = O(0.1 - 1)$ is likely to be the most physically relevant regime. Previous studies \citep{CROWETAYLOR3} have noted that TTW balance can remain valid even for the case of finite Rossby numbers so numerical simulations are now performed to test if the phenomenon described above are relevant to this regime.

\subsection{Numerical simulations for Ro = 1}

Here, \cref{eq:TTW} is solved subject to no-stress and no-flux boundary conditions using the Dedalus package \citep{BurnsVOLB20}. These numerical simulations are two dimensional; the cross-front direction is taken to align with the $y$ axis where non-traditional effects are maximised and $\pderline{}{x}$ is set to zero. Fields are expanded in terms of a Fourier basis in the horizontal $(y)$ direction and a Chebyshev basis in the vertical $(z)$ direction and time-stepped using a third order implicit-explicit Runge-Kutta scheme. Horizontal mixing with a viscosity of $10^{-4}$ is included for numerical stability. Simulations are initialised using the $O(1)$ TTW solution for velocity and buoyancy given in \cref{sec:summ} for $b_0 = \tanh y$ and $\psi_0 = 0$.

\begin{figure}
	\centering
	\begin{subfigure}[b]{0.485\textwidth}
	\centering
	\includegraphics[trim={0 0 0 0},clip,width=\textwidth]{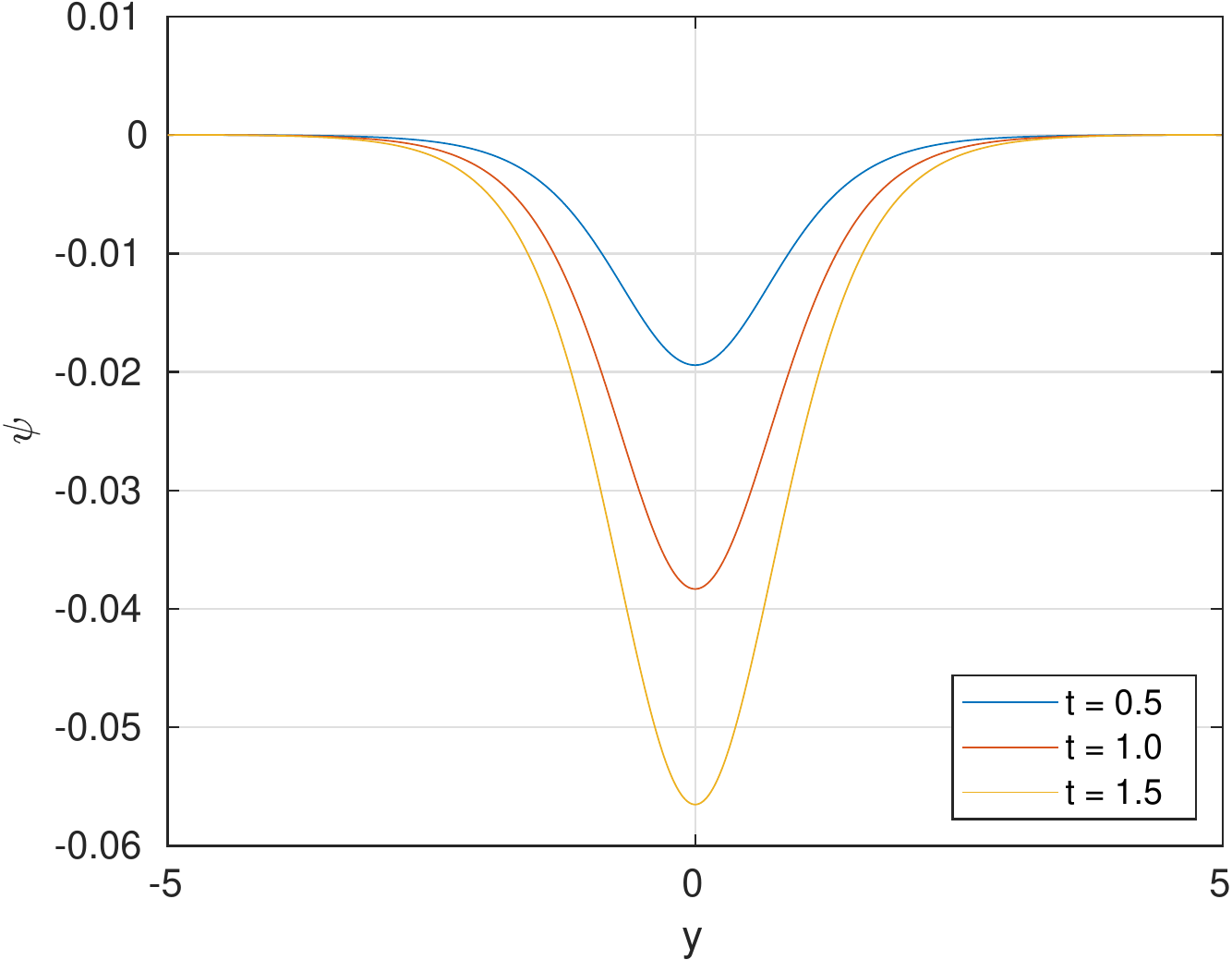}
	\caption{}
	\end{subfigure}
	\begin{subfigure}[b]{0.495\textwidth}
	\centering
	\includegraphics[trim={0 0 0 0},clip,width=\textwidth]{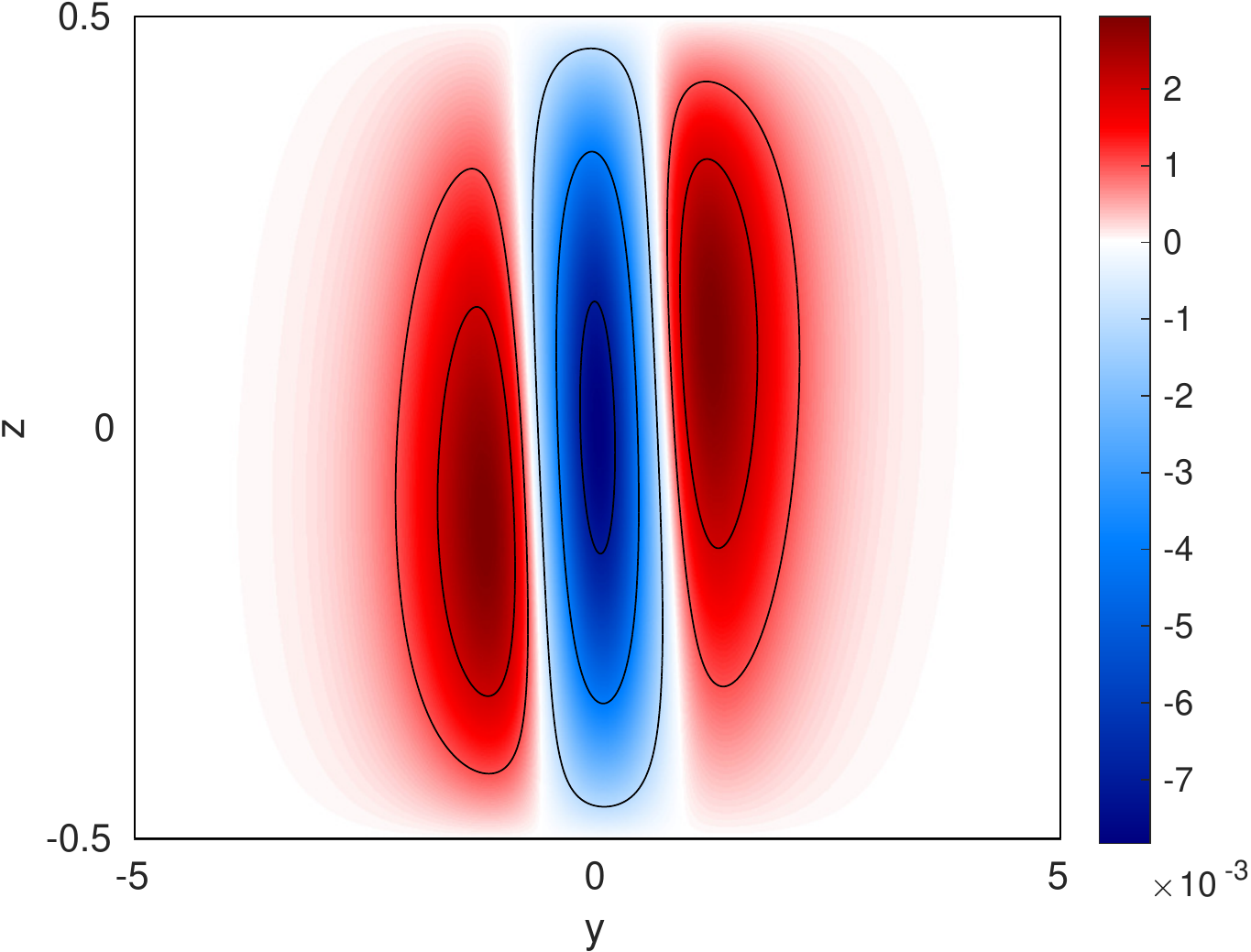}
	\caption{}
	\end{subfigure}
	\caption{Results from a numerical simulation with $(\Ro,\delta,\E) = (0.1,1,0.1)$ showing (a) the streamfunction, $\psi$, describing the depth-averaged along-front flow and (b) the circulation component $\phi_1$ at $t = 2$.}
	\label{fig:num_sol_1}
\end{figure}

\cref{fig:num_sol_1} shows numerical results for $\Ro = 0.1$, $\delta = 1$ and $\E = 0.1$. \cref{fig:num_sol_1}.(a) shows the development of the along-front jets through the growth of streamfunction of the  depth-averaged flow, $\psi$, with time. These profiles for $\psi$ are consistent with the analytical predictions shown in \cref{fig:psi}.(a), differing by less than $3\%$ from the theory despite the use of an order $1$ value of $\delta$ in an asymptotic expression that is known only to $O(\delta)$. 
\cref{fig:num_sol_1}.(b) shows the $O(\delta)$ component of the streamfunction of the frontal circulation, $\phi_1$, at $t = 2$. Here $\phi_1$ is calculated as the difference between the total value of $\phi$ and the value of $\phi_0$ calculated using \cref{eq:phi_sol}. Again the results are well described by the theory as the structure of these circulation cells can be seen to be a sum of the components shown in \cref{fig:phi_v}.(a) and \cref{fig:phi_v}.(b). The accuracy of the theoretical predictions for order $1$ values of $\delta$ suggests that the solutions of \cref{sec:summ} are valid, even outside of the asymptotic regime considered.

\begin{figure}
	\centering
	\begin{subfigure}[b]{0.485\textwidth}
	\centering
	\includegraphics[trim={0 0 0 0},clip,width=\textwidth]{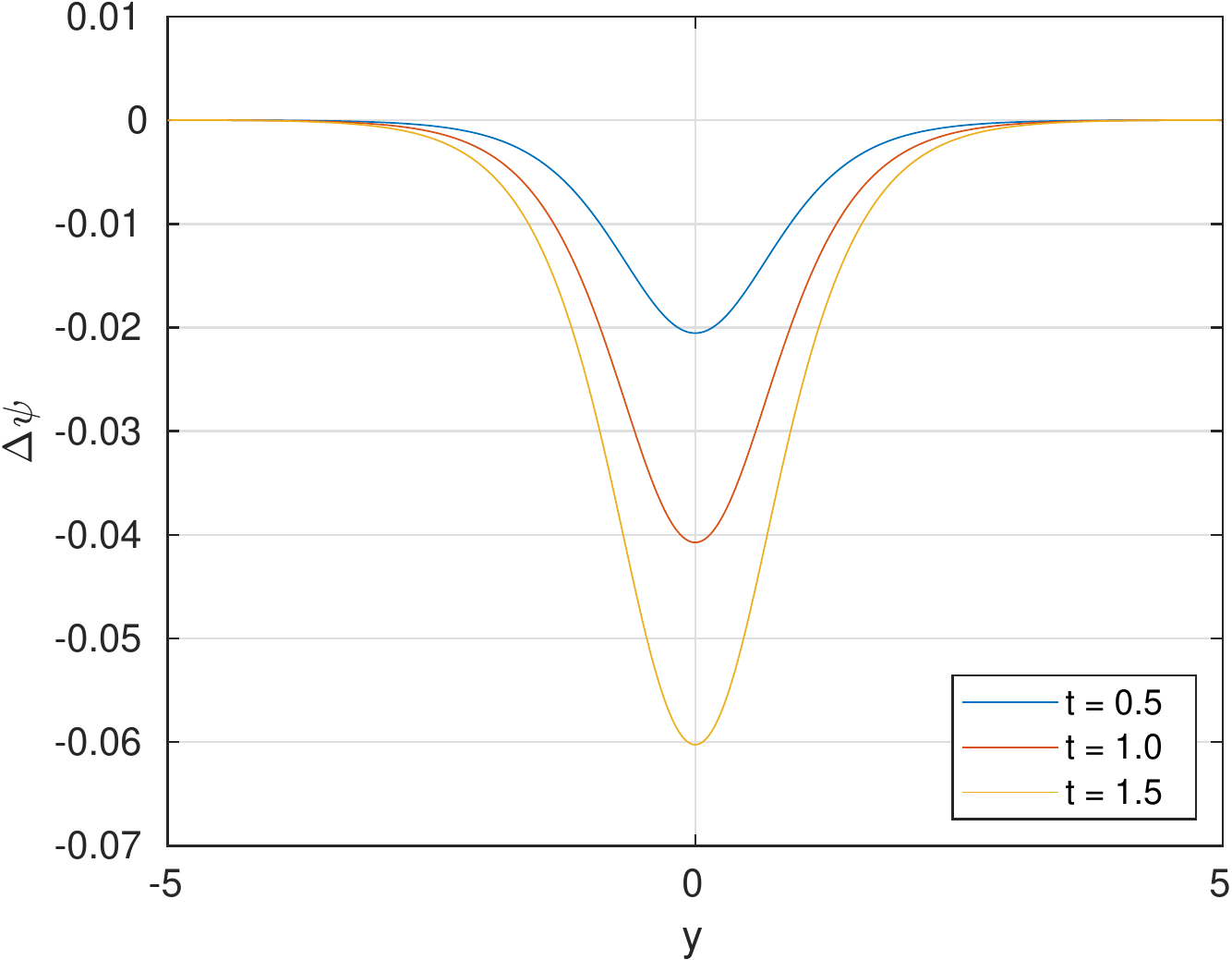}
	\caption{}
	\end{subfigure}
	\begin{subfigure}[b]{0.495\textwidth}
	\centering
	\includegraphics[trim={0 0 0 0},clip,width=\textwidth]{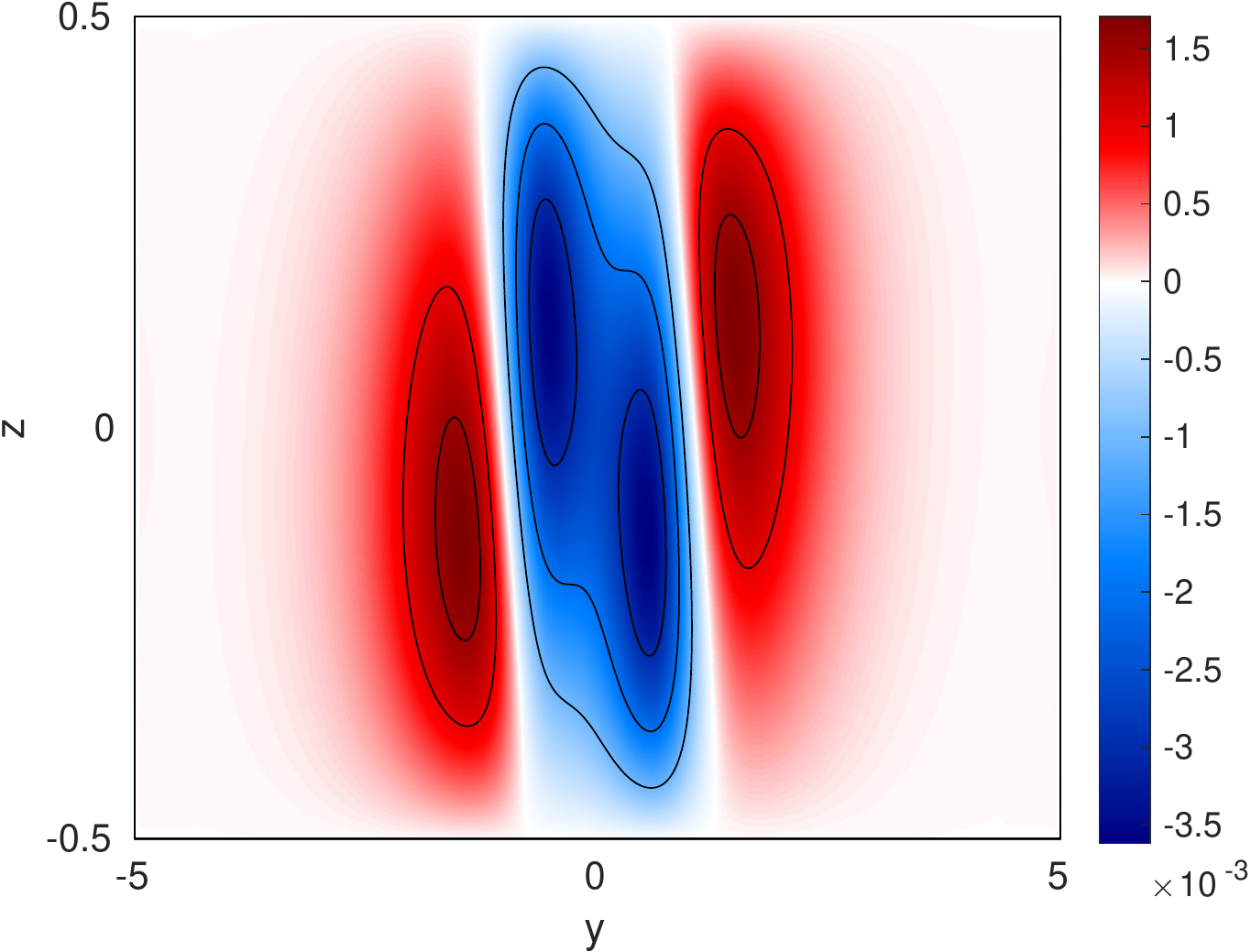}
	\caption{}
	\end{subfigure}
	\caption{Results from a numerical simulation with $(\Ro,\delta,\E) = (1,1,0.1)$ showing (a) the difference in streamfunction, $\Delta\psi$, between the results for $\delta = 1$ and $\delta = 0$ and (b) the difference in frontal circulation, $\Delta\phi,$ at $t = 2$.}
	\label{fig:num_sol_2}
\end{figure}

\cref{fig:num_sol_2} shows numerical results for $\Ro = 1$, $\delta = 1$ and $\E = 0.1$.  Here, the effects of nonlinearity become significant and it is necessary to separate the effects of finite $\Ro$ from the effects of finite $\delta$. This may be done by running another simulation with $(\Ro,\delta,\E) = (1,0,0.2)$ and calculating the difference
\begin{equation}
\Delta \varphi = \varphi|_{\delta = 1} - \varphi|_{\delta = 0},
\end{equation}
for some field $\varphi$. \cref{fig:num_sol_2}.(a) shows the value of $\Delta \psi$ for a range of value of $t$. Similarly to the case of $\Ro = 0.1$, the along front flow is found to be well described by the theoretical predictions with a difference of around $4\%$ between $\Delta \psi$ and the prediction for $\psi_0+\delta\,\psi_1$. The difference in frontal circulation, $\Delta\phi$, is shown in \cref{fig:num_sol_2}.(b). While qualitatively similar to theoretical predictions and the case of $\Ro = 0.1$, the nonlinearity and non-traditional components appear to interact nonlinearly resulting in some deviation from the predictions. In particular, the centre of the middle circulation cell in \cref{fig:num_sol_2}.(b) is seen to split in two. Nonetheless, even for cases far outside the asymptotic regime considered analytically, the effects of including non-traditional rotation appear to be qualitatively the same as discussed above, with the generation of along-front jets and a modification of the frontal circulation.

\section{Discussion and conclusions}
\label{sec:diss}

Here the effects of the non-traditional component of rotation on a front in turbulent thermal wind balance have been considered. Solutions are calculated as a perturbation of the TTW solutions of \citet{CROWETAYLOR,CROWETAYLOR3} using an asymptotic approach. The magnitude of the non-traditional correction terms is found to depend strongly on the direction of the front. Fronts where the along-front direction is aligned with North-South are found to be unaffected by the non-traditional rotation terms. Conversely, non-traditional effects are maximised for fronts aligned with East-West.

A primary effect of the non-traditional rotation is the generation of vertical vorticity by the horizontal component of the non-traditional Coriolis force, $\tw{f}w$. This vorticity is generated in regions of strong vertical velocity and manifests as along-front jets. Over timescales of $t \sim 1/(\delta f)$ strong vorticity generation is expected, resulting in a modification of the total vertical vorticity of the system once the generated vorticity is of similar magnitude to the planetary vorticity, $f$. In this case, relative vorticity terms must be included in the leading order balance \citep{WENEGRAT_THOMAS,CROWETAYLOR3} resulting in a modified leading order solution.

Additionally, the vertical component of the non-traditional Coriolis force, $-\tw{f}u$, acts to drive the system out of hydrostatic balance resulting in a new pressure component and hence a new horizontal pressure gradient. Since the velocity, $u$, may be split into a component corresponding to the background vorticity and a component corresponding to the TTW flow, two new velocity contributions are obtained. Firstly, the background vorticity drives a modification to the leading order TTW flow by changing the horizontal pressure gradients. Secondly, the action of both non-traditional components of the Coriolis force on the leading order TTW solution drives a small correction flow consisting of several circulation cells. The combined effect of these contributions may lead to a change in the topographic structure of the total frontal circulation, which, for vanishing non-traditional effects, consists of a single cell. Further, the modification of this circulation may act to enhance the exchange of tracers through the mixed layer.

As observed in \citet{CROWETAYLOR}, the TTW velocity field consists of a leading-order circulation around the front. This circulation acts to re-stratify the front and the stratification is maintained through an advection-diffusion balance in the buoyancy equation. Since non-traditional effects modify this circulation, the stratification is modified by the appearance of terms which depends both on the background buoyancy gradient and the background vorticity. Some circulation components are observed to be frontogenetic, driving a sharpening of horizontal buoyancy gradients. However, outside of a small region around the equator where the analysis is not valid, this frontogenesis is expected to be weak when compared with other mechanisms \citep{HOSKINSBRETHERTON,SHAKESPEARETAYLOR,MCWILLIAMS}.

The correlation between the cross-front flow and the vertical buoyancy gradient may drive the evolution of the background buoyancy field through shear dispersion. Non-traditional effects are expected to affect this process predominantly via the modification of the velocity and buoyancy fields by the generated background vorticity. It should be noted that an important feature of fronts is the presence of baroclinic instability \citep{STONE1966} which can also modify the background buoyancy field. Since baroclininc instability would be expected to act over faster time scales than shear dispersive spreading and can exist in the presence of strong vertical mixing \citep{CROWETAYLOR2}, these instabilities should be considered when studying the long-term behaviour of the front.

Using typical frontal parameters of $H = 100\,\textrm{m}$ and $L = 10 \,\textrm{km}$, the value of $\delta$ is expected to be very small ($\delta \leq 0.01$) away from the tropical and subtropical regions. Therefore, the non-traditional component of rotation is unlikely to play a significant role in general frontal dynamics. However, in the low latitude regions near the equator it may be possible to get $\delta \sim 0.1\, -\, 1$ so fronts in these regions may have dynamics which are strongly affected by non-traditional effects. An order $1$ value of $\delta$ requires a fairly small frontal width of $L \sim 1\,\textrm{km}$ so fronts in this regime are also expected to have order $1$ values of the Rossby number with nonlinear advection playing an important role. While the asymptotic results presented in \cref{sec:summ} are not strictly valid outside of the regime $\Ro\ll\delta\ll 1$, numerical simulations indicate that the same phenomenon occur and that these solutions can provide accurate predictions for the case of finite Rossby numbers and finite non-traditional parameters even if they are not formally valid.

Another limitation of the asymptotic model is the idealised setup with turbulent mixing represented by a constant turbulent Ekman number and any large-scale geostrophic flow components being neglected. The inclusion of a more realistic turbulence parametrisation and a background flow field require a numerical approach and is a topic for future work.

\bigskip
\noindent{\bf Acknowledgements\bf{.}} The author would like to thank Dr. John Taylor for helpful comments on an early draft of the paper and three anonymous reviewers whose constructive comments have greatly improved this manuscript.\\

\noindent{\bf Declaration of Interests\bf{.}} The author reports no conflict of interest.

\appendix

\section{The O($\delta$) depth-dependent velocity}
\label{sec:app_sol1}

Here \cref{eq:TTW1_step1} is solved for the depth-dependent component on the $O(\delta)$ velocity field. Substituting for $w_0'$ using \cref{eq:TTW0_w} and $p_1'$ using \cref{eq:p_1} gives
\begin{subequations}
\label{eq:TTW1_step2}
\begin{alignat}{3}
v_1+\pder{^2u_1}{\z^2} = & \,\E\pder{}{y}\left[K_1(\z) \pder{b_0}{y} -K_2(\z)\pder{b_0}{x}\right]+z\left[\pder{b_1}{x}+\pder{\da{u}_0}{x}\right],\\
u_1-\pder{^2v_1}{\z^2} = & \,\E\pder{}{y}\left[K_1(\z)\pder{b_0}{x} +K_2(\z)\pder{b_0}{y}\right]-z\left[\pder{b_1}{y}+\pder{\da{u}_0}{y}\right],
\end{alignat}
\end{subequations}
where
\begin{equation}
K_1(\z) = K'(\z)-\frac{K(\z_0)}{\z_0}, \quad K_2(\z) = K'''(\z)-\frac{K''(\z_0)}{\z_0}+\frac{\z^2}{2}-\frac{\z_0^2}{6},
\end{equation}
and $\z = z/\sqrt{\E}$ as before. The right-hand sides of \cref{eq:TTW1_step2} consist of two forcing terms in square brackets, these can now be treated separately by linearity and a superscript ($\{1\}$ and $\{2\}$) will be used to denote which forcing term a solution corresponds to. The second forcing term resembles that of the leading order system, $-\nabla_H p_0 = -z \nabla_H b_0$, so can be solved similarly for solution
\begin{subequations}
\label{eq:TTW_sol_1_part_1}
\begin{alignat}{3}
u_1'^{\{2\}} = & -\sqrt{\E}\left[K''(\z)\pder{}{x}-K(\z)\pder{}{y}\right]\left(b_1-\pder{\psi_0}{y}\right),\\
v_1'^{\{2\}} = & -\sqrt{\E}\left[K(\z)\pder{}{x}+K''(\z)\pder{}{y}\right]\left(b_1-\pder{\psi_0}{y}\right),
\end{alignat}
\end{subequations}
where $\da{u}_0$ has been replaced using $\da{u}_0 = -\pderline{\psi_0}{y}$. The first forcing term is more complicated but the system may be solved by taking
\begin{subequations}
\begin{alignat}{3}
u_1'^{\{1\}} = & \E\pder{}{y}\left[A(\z)\pder{b_0}{x}-B(\z)\pder{b_0}{y}\right],\\
v_1'^{\{1\}} = & \E\pder{}{y}\left[B(\z)\pder{b_0}{x}+A(\z)\pder{b_0}{y}\right],
\end{alignat}
\end{subequations}
based on the form of the equations. The functions $A(\z)$ and $B(\z)$ satisfy
\begin{equation}
\label{eq:AB_eqn}
A-B'' = K'(\z)-\frac{K(\z_0)}{\z_0}, \quad \textrm{and} \quad -B-A'' = K'''(\z)-\frac{K''(\z_0)}{\z_0}+\frac{\z^2}{2}-\frac{\z_0^2}{6},
\end{equation}
which may be solved with no-stress boundary conditions to obtain solutions for $A$ and $B$. The solutions for each forcing term may now be summed to give the final solution for $(u_1',v_1')$.

\section{Vertical structure functions}
\label{sec:app1}

The vertical structure functions, $A(\z)$ and $B(\z)$, are determined as solutions of \cref{eq:AB_eqn}. The first rows of the following solutions give the particular solution required to solve \cref{eq:AB_eqn} while the second rows give the complementary function component required to satisfy no-stress boundary conditions on the top and bottom boundary. Solutions are
\begin{equation}
\begin{gathered}
A(\z) = -\frac{\z K''(\z)+4}{2} - \frac{K(\z_0)}{\z_0} + \hspace{180pt}\\ \frac{\left(4\z_0^2+5\z_0 K(\z_0)+(K(\z_0))^2+(K''(\z_0))^2\right)(K'(\z)+1)+3\z_0K''(\z_0)K'''(\z)}{2\left[(K''(\z_0))^2+(K(\z_0)+\z_0)^2\right]},
\end{gathered}
\end{equation}
and
\begin{equation}
\begin{gathered}
B(\z) = -\frac{\z K(\z)+2\z^2}{2} + \frac{K''(\z_0)}{\z_0}+\frac{\z_0^2}{6} + \hspace{150pt}\\- \frac{\left(4\z_0^2+5\z_0 K(\z_0)+(K(\z_0))^2+(K''(\z_0))^2\right)K'''(\z)-3\z_0K''(\z_0)\left(K'(\z)+1\right)}{2\left[(K''(\z_0))^2+(K(\z_0)+\z_0)^2\right]}.
\end{gathered}
\end{equation}
The function $C(\z)$ describes the vertical velocity and is calculated as a single vertical integral of $A$ by mass conservation. The integration constant is taken such that $C$ is zero on the top and bottom boundaries so $C$ is given by
\begin{equation}
\begin{gathered}
C(\z) = 2(\z-\z_0)+\frac{1}{2}(K(\z)+K(\z_0))-\frac{\z K'(\z)}{2}-\frac{K(\z_0)\z}{\z_0}  \hspace{80pt}\\ + \frac{\left(4\z_0^2+5\z_0 K(\z_0)+(K(\z_0))^2+(K''(\z_0))^2\right)(K(\z)+\z-K(\z_0)-\z_0)}{2\left[(K''(\z_0))^2+(K(\z_0)+\z_0)^2\right]}+\\ \hspace{130pt}\frac{3\z_0K''(\z_0)(K''(\z)-K''(\z_0))}{2\left[(K''(\z_0))^2+(K(\z_0)+\z_0)^2\right]}.
\end{gathered}
\end{equation}

The structure functions which determine the structure of the vertical stratification are $D_1(\z)$ and $D_2(\z)$which are determined as solutions of $D_1''(\z) = A(\z)$ and $D_2''(\z) = B(\z)$ with boundary conditions of no flux on the top and bottom boundaries (corresponding to a vanishing first derivative on $\z=\pm\z_0$). Solutions are
\begin{subequations}
\begin{alignat}{3}
D_1(\z) = & B(\z)-\left[K'''(\z)-\frac{K''(\z_0)}{\z_0}\right] - \left(\frac{K(\z_0)}{2\z_0}+\frac{1}{2}\right)\left(\z^2-\frac{1}{3}\z_0^2\right),\\
D_2(\z) =& -A(\z)-\left[K'(\z)-\frac{K(\z_0)}{\z_0}\right] + \left(\frac{K''(\z_0)}{2\z_0}+\frac{\z_0^2}{12}\right)\left(\z^2-\frac{1}{3}\z_0^2\right) - \frac{1}{24}\left(\z^4-\frac{1}{5}\z_0^4\right).
\end{alignat}
\end{subequations}

\bibliographystyle{jfm}
\bibliography{bibliography}

\begin{thebibliography}{32}
\expandafter\ifx\csname natexlab\endcsname\relax\def\natexlab#1{#1}\fi

\bibitem[Blumen(2000)]{BLUMEN}
{\sc Blumen, W.} 2000 Inertial oscillations and frontogenesis in a zero
  potential vorticity model. {\em J. Phys. Oceanogr.\/} {\bf 30}, 31--39.

\bibitem[Burns {\em et~al.\/}(2020)Burns, Vasil, Oishi, Lecoanet \&
  Brown]{BurnsVOLB20}
{\sc Burns, K.~J., Vasil, G.~M., Oishi, J.~S., Lecoanet, D. \& Brown, B.~P.}
  2020 Dedalus: {A} flexible framework for numerical simulations with spectral
  methods. {\em Phys. Rev. Res.\/} {\bf 2}, 023068.

\bibitem[Charney(1973)]{CHARNEY}
{\sc Charney, J.~G.} 1973 {\em Planetary Fluid Dynamics\/}, chap. Symmetric
  Circulations in Idealized Models, pp. 128--141. D. Reidel Publishing Company.

\bibitem[Coleman {\em et~al.\/}(1990)Coleman, Ferziger \& Spalart]{colemanetal}
{\sc Coleman, G.~N., Ferziger, J.~H. \& Spalart, P.~R.} 1990 A numerical study
  of the turbulent ekman layer. {\em J. Fluid Mech.\/} {\bf 213}, 313–348.

\bibitem[Cronin \& Kessler(2009)]{CRONINKESSLER}
{\sc Cronin, Meghan~F \& Kessler, William~S} 2009 Near-surface shear flow in
  the tropical {Pacific} cold tongue front. {\em J. Phys. Oceanogr.\/} {\bf
  39}~(5), 1200--1215.

\bibitem[Crowe \& Taylor(2018)]{CROWETAYLOR}
{\sc Crowe, M.~N. \& Taylor, J.~R.} 2018 The evolution of a front in turbulent
  thermal wind balance, part 1. {T}heory. {\em J. Fluid Mech.\/} {\bf 850},
  179--211.

\bibitem[Crowe \& Taylor(2019{\natexlab{{\em a\/}}})]{CROWETAYLOR2}
{\sc Crowe, M.~N. \& Taylor, J.~R.} 2019{\natexlab{{\em a\/}}} Baroclinic
  instability with a simple model for vertical mixing. {\em J. Phys.
  Oceanogr.\/} {\bf 49}, 3273--3300.

\bibitem[Crowe \& Taylor(2019{\natexlab{{\em b\/}}})]{CROWETAYLOR3}
{\sc Crowe, M.~N. \& Taylor, J.~R.} 2019{\natexlab{{\em b\/}}} The evolution of
  a front in turbulent thermal wind balance, part 2. {N}umerical simulations.
  {\em J. Fluid Mech.\/} {\bf 880}, 326--352.

\bibitem[Crowe \& Taylor(2020)]{CROWETAYLOR20}
{\sc Crowe, M.~N. \& Taylor, J.~R.} 2020 The effects of surface wind stress and
  buoyancy flux on the evolution of a front in a turbulent thermal wind
  balance. {\em Fluids\/} {\bf 5}~(2).

\bibitem[Eckart(1960)]{ECKART}
{\sc Eckart, C.} 1960 {\em Hydrodynamics of Oceans and Atmospheres\/}.
  Pergamon, 290 pp.

\bibitem[Eliassen(1962)]{ELIASSEN}
{\sc Eliassen, A.} 1962 On the vertical circulation in frontal zones. {\em
  Geofys. Publ.\/} {\bf 24}~(4), 147--160.

\bibitem[Ferrari(2011)]{FERRARI}
{\sc Ferrari, R.} 2011 A frontal challenge for climate models. {\em Science\/}
  {\bf 332}~(6027), 316--317.

\bibitem[Garrett \& Loder(1981)]{GARRETTLODER}
{\sc Garrett, C. J.~R. \& Loder, J.~W.} 1981 Dynamical aspects of shallow sea
  fronts. {\em Phil. Trans. R. Soc. Lond. A\/} {\bf 302}, 563--581.

\bibitem[Garwood(1991)]{GARWOOD}
{\sc Garwood, R.~W.} 1991 Enhancements to deep turbulent entrainment. In {\em
  Deep Convection and Deep Water Formation in the Oceans\/} (ed. P.C. Chu \&
  J.C. Gascard), {\em Elsevier Oceanography Series\/}, vol.~57, pp. 197--213.
  Elsevier.

\bibitem[Gerkema(2006)]{GERKEMA_2006}
{\sc Gerkema, T.} 2006 {Internal-wave reflection from uniform slopes: higher
  harmonics and {C}oriolis effects}. {\em Nonlin. Proc. Geophys.\/} {\bf
  13}~(3), 265--273.

\bibitem[Gerkema \& Shira(2005)]{gerkema_shrira_2005}
{\sc Gerkema, T. \& Shira, V.~I.} 2005 Near-inertial waves in the ocean: beyond
  the ‘traditional approximation’. {\em Journal of Fluid Mechanics\/} {\bf
  529}, 195–219.

\bibitem[Gerkema {\em et~al.\/}(2008)Gerkema, Zimmerman, Maas \& van
  Haren]{GERKEMAETAL}
{\sc Gerkema, T., Zimmerman, J. T.~F., Maas, L. R.~M. \& van Haren, H.} 2008
  Geophysical and astrophysical fluid dynamics beyond the traditional
  approximation. {\em Rev. Geophys.\/} {\bf 46}~(2).

\bibitem[Gula {\em et~al.\/}(2014)Gula, Molemaker \& McWilliams]{GULAETAL}
{\sc Gula, J., Molemaker, M.~J. \& McWilliams, J.~C.} 2014 Submesoscale cold
  filaments in the {Gulf Stream}. {\em J. Phys. Oceanogr.\/} {\bf 44},
  2617--2643.

\bibitem[Hoskins(1982)]{HOSKINS}
{\sc Hoskins, B.~J.} 1982 The mathematical theory of frontogenesis. {\em Annu.
  Rev. Fluid Mech.\/} {\bf 14}, 131--151.

\bibitem[Hoskins \& Bretherton(1972)]{HOSKINSBRETHERTON}
{\sc Hoskins, B.~J. \& Bretherton, F.~P.} 1972 Atmospheric frontogenesis
  models: Mathematical formulation and solution. {\em J. Atmos. Sci.\/} {\bf
  29}, 11--37.

\bibitem[Hua {\em et~al.\/}(1997)Hua, Moore \& Gentil]{hua_etal}
{\sc Hua, B.~L., Moore, D.~W. \& Gentil, S.~L.} 1997 Inertial nonlinear
  equilibration of equatorial flows. {\em J. Fluid Mech.\/} {\bf 331},
  345–371.

\bibitem[Lucas {\em et~al.\/}(2017)Lucas, McWilliams \& Rousseau]{LUCAS_ETAL}
{\sc Lucas, C., McWilliams, J.~C. \& Rousseau, A.} 2017 Large scale ocean
  models beyond the traditional approximation. {\em Annales de la Facult{\'{e}}
  des sciences de Toulouse : Math{\'{e}}matiques\/} {\bf Ser. 6, 26}~(4),
  1029--1049.

\bibitem[McWilliams(2017)]{MCWILLIAMS}
{\sc McWilliams, J.~C.} 2017 Submesoscale surface fronts and filaments:
  secondary circulation, buoyancy flux, and frontogenesis. {\em J. Fluid
  Mech.\/} {\bf 823}, 391--432.

\bibitem[McWilliams {\em et~al.\/}(2015)McWilliams, Gula, Molemaker, Renault \&
  Shchepetkin]{MCWILLIAMSETAL}
{\sc McWilliams, J.~C., Gula, J., Molemaker, M.~J., Renault, L. \& Shchepetkin,
  A.~F.} 2015 Filament frontogenesis by boundary layer turbulence. {\em J.
  Phys. Oceanogr.\/} {\bf 45}, 1988--2005.

\bibitem[McWilliams \& Huckle(2006)]{MCWILLIAMSHUCKLE}
{\sc McWilliams, J.~C. \& Huckle, E.} 2006 Ekman layer rectification. {\em J.
  Phys. Oceanogr.\/} {\bf 36}~(8), 1646 -- 1659.

\bibitem[Orlanski \& Ross(1977)]{ORLANSKIROSS}
{\sc Orlanski, I. \& Ross, B.~B.} 1977 The circulation associated with a cold
  front: Part i: Dry case. {\em J. Atmos. Sci.\/} {\bf 34}, 1619--1633.

\bibitem[Shakespeare \& Taylor(2013)]{SHAKESPEARETAYLOR}
{\sc Shakespeare, C.~J. \& Taylor, J.R.} 2013 A generalized mathematical model
  of geostrophic adjustment and frontogenesis: uniform potential vorticity.
  {\em J. Fluid Mech.\/} {\bf 736}, 366--413.

\bibitem[Sheremet(2004)]{sheremet_2004}
{\sc Sheremet, V.~A.} 2004 Laboratory experiments with tilted convective plumes
  on a centrifuge: a finite angle between the buoyancy force and the axis of
  rotation. {\em J. Fluid Mech.\/} {\bf 506}, 217–244.

\bibitem[Stone(1966)]{STONE1966}
{\sc Stone, P.~H.} 1966 On non-geostrophic baroclinic stability. {\em J. Atmos.
  Sci.\/} {\bf 23}, 390--400.

\bibitem[de~Verdi{\`{e}}re \& Schopp(1994)]{verdiere_schopp_1994}
{\sc de~Verdi{\`{e}}re, A.~Colin \& Schopp, R.} 1994 Flows in a rotating
  spherical shell: the equatorial case. {\em J. Fluid Mech.\/} {\bf 276},
  233–260.

\bibitem[Wenegrat \& McPhaden(2016)]{WENEGRAT}
{\sc Wenegrat, J.~O. \& McPhaden, M.~J.} 2016 Wind, waves, and fronts:
  Frictional effects in a generalized ekman model. {\em J. Phys. Oceanogr.\/}
  {\bf 46}~(2), 371--394.

\bibitem[Wenegrat \& Thomas(2017)]{WENEGRAT_THOMAS}
{\sc Wenegrat, J.~O. \& Thomas, L.~N.} 2017 Ekman transport in balanced
  currents with curvature. {\em J. Phys. Oceanogr.\/} {\bf 47}~(5), 1189 --
  1203.

\end{thebibliography}

\end{document}